\documentclass{article}

\usepackage{PRIMEarxiv}

\usepackage[utf8]{inputenc} 
\usepackage[T1]{fontenc}    
\usepackage{hyperref}       
\usepackage{url}            
\usepackage{booktabs}       
\usepackage{amsfonts}       
\usepackage{nicefrac}       
\usepackage{microtype}      
\usepackage{lipsum}
\usepackage{fancyhdr}       
\usepackage{graphicx}       
\graphicspath{{imgs/}}     
\usepackage{authblk}
\usepackage{amsmath}
\usepackage{optidef}
\usepackage{enumitem}
\usepackage{caption}
\usepackage{subcaption}
\usepackage{url}
\usepackage{mdframed}

\usepackage{cite}
\usepackage{amsmath,amssymb,amsfonts}
\usepackage{algorithmic}
\usepackage{graphicx}
\usepackage{longtable}
\usepackage{multirow}
 \usepackage{url}
\usepackage{textcomp}
\usepackage{braket}
\usepackage{subcaption}

\usepackage{array}

\begingroup
\catcode`\_=\active
\gdef\mathurl#1{\text{\url{#1}}}
\endgroup

\let\oldurl\url
\renewcommand{\url}[1]{\text{\oldurl{#1}}}

\title{Quantum Reservoir Computing for Credit Card Default Prediction on a Neutral Atom Platform
}

\author[1,2]{Giacomo Vitali}
\author[1,2]{Chiara Vercellino}
\author[1]{Paolo Viviani}
\author[1]{Olivier Terzo}
\author[1]{Bartolomeo Montrucchio}
\author[3]{Valeria Zaffaroni}
\author[3]{Francesca Cibrario}
\author[3]{Christian Mattia}
\author[3]{Giacomo Ranieri}
\author[3]{Alessandro Sabatino}
\author[3]{Francesco Bonazzi}
\author[3]{Davide Corbelletto}
\affil[1]{Advanced Computing, Photonics and Electromagnetics Research Domain, Fondazione LINKS, Torino 10138, Italy\\
\texttt{firstname.lastname@linksfoundation.com}}
\affil[2]{Control And Computer Engineering Department (DAUIN), Politecnico di Torino, Torino 10129, Italy\\
\texttt{firstname.lastname@polito.it}}
\affil[3]{Intesa Sanpaolo, Torino 10121, Italy\\
\texttt{firstname.lastname@intesasanpaolo.com}}

\begin{document}

\maketitle

\begin{mdframed}[linewidth=0.5pt]
This work has been submitted to the IEEE for possible publication. 
Copyright may be transferred without notice, after which this version 
may no longer be accessible.
\end{mdframed}
\vspace{1cm}

\begin{abstract}
In this paper, we define and benchmark a hybrid quantum-classical machine learning pipeline by performing a binary classification task applied to a real-world financial use case. Specifically, we implement a Quantum Reservoir Computing (QRC) layer within a classical routine that includes data preprocessing and binary classification. The reservoir layer has been executed on \textit{QuEra}'s \textit{Aquila}, a 256-qubit neutral atom simulator, using two different types of encoding: position and local detuning. In the former case, classical data are encoded into the relative distance between atoms; in the latter, into pulse amplitudes. The developed pipeline is applied to predict credit card defaults using a public dataset and a wide variety of traditional classifiers. The results are compared with a fully-classical pipeline including a Deep Neural Network (DNN) model.\\ Additionally, the impact of hardware noise on classification performance is evaluated by comparing the results obtained using Aquila within the classification workflow with those  obtained using a classical, noiseless emulation of the quantum system.
The results indicate that the noiseless emulation achieves competitive performance with the fully-classical pipeline, while noise significantly degrades overall performance.
Although the results for this specific use case are comparable to those of the classical benchmark, the flexibility and scalability of QRC highlight strong potential for a wide range of applications.
\end{abstract}


\section{Introduction}\label{sec:introduction}

The field of Quantum Machine Learning (QML) has been theoretically explored since the early era of Quantum Computing (QC). Although several architectures have been proposed over the years~\cite{PERALGARCIA2024100619}, the first examples are those that require features (such as QRAM and fault-tolerance) which are still not available on state-of-the-art quantum computers. In fact, the Noisy Intermediate-Scale Quantum (NISQ)-era platforms are characterized by a limited number of qubits (in the order of a few hundreds at most and a few thousands in the case of annealers) and by short coherence time. For this reason, a second generation of QML methods for NISQ quantum computers have been designed to have low-depth quantum circuits and smaller number of qubits. Some of these NISQ-ready QML algorithms have been tested on mock or reduced data~\cite{simoes2023experimental, batra2021quantum}, but encoding real data into the quantum register is still a hard task. Moreover, most of the NISQ-ready methods are variational, which makes any potential computational advantage difficult to prove.
Nonetheless, the emergence of analog quantum simulators bears the promise of more efficient execution of some classes of NISQ methodologies. However, such machines are considered a niche, with few examples of specifically designed QML methods~\cite{PhysRevA.107.042615, PhysRevA.104.032416}.

In this paper, we present the implementation and execution of a hybrid classification technique, which includes a Quantum Reservoir Computing (QRC) layer, to a real-world problem to probe the limits and applicability of NISQ-era analog quantum computers. To this end, we decided to target the analog neutral atom quantum simulator \textit{Aquila}~\cite{wurtz2023aquilaqueras256qubitneutralatom}, built by \textit{QuEra}. This platform, in fact, offers 256 qubits, the highest for any publicly available quantum simulator, that can be distributed over a large registry area, and most importantly, the possibility to apply local detuning to the atom arrays. To the best of our knowledge, \textit{Aquila} is the only platform that provides access to the latter (although as an experimental feature at the time we conducted our experiments) among cloud-accessible neutral atom simulators.

In particular, \textit{Aquila}'s effective Hamiltonian (considering $n$ qubits) takes the form:

\begin{align}\label{hamiltonian_eq}
    \hat{\mathcal{H}}(t) = & \frac{\hslash \Omega(t)}{2} \sum_{i=1}^{n} ( e^{i\phi(t)} \ket{0_i}\bra{1_i} + e^{-i\phi(t)} \ket{1_i}\bra{0_i} ) \nonumber\\ 
    & - \sum_{i=1}^{n} \hslash \Delta_i(t) \hat{n}_i + \sum_{j<i} \frac{C_6}{ r_{i,j} ^{6}} \hat{n}_i \hat{n}_j ,
\end{align}

where $\Omega(t)$ is the time-dependent global Rabi drive, $\phi(t)$ its relative phase, $\Delta_i(t)$ the site-dependent (or local) detuning, and $\hat{n}_i$ is the operator $\ket{1_i}\bra{1_i}$ that counts the atoms in the excited state and $C_6$ is the Rydberg interaction coefficient, whose value depends on the chosen Rydberg level. The last term, the Van der Waals interaction, is often rewritten using $V_{ij} = \frac{C_6}{ r_{i,j} ^{6}}$, where $r_{i,j}$ is the relative distance between the $i^{th}$ and $j^{th}$ atoms.\\
For a given dataset and problem, the input features of the QRC layer must be encoded into the quantum system by mapping them to the Hamiltonian parameters. These parameters are used to evolve the system over time. The output features are built by defining a set of observable operators which are then measured for a predetermined set of timesteps to capture the time dependence of the original input features. Physically, this can be implemented through weak measurements~\cite{Yasuda:2023vop} or by performing a Hamiltonian simulation for each timestep, depending on the characteristic of the target platform. Finally, the output features of the QRC layer are fed into a classical classifier for the final prediction.

We decided to test the effectiveness of the methodology in the context of an imbalanced classification problem. We choose one of such problems from the financial sector, namely the prediction of default for credit card owners.
The purpose of this prediction is to determine whether customers can settle their credit card balance in the following month. The problem therefore is highly imbalanced since the number of defaults is smaller than the number of non-defaults.
For this specific real-world example, to ensure the reproducibility of our experiments, we opted for a publicly available dataset~\cite{default_of_credit_card_clients_350} that has been frequently referenced in prior studies~\cite{YEH20092473, islam2018creditdefaultminingusing, XIONG2013665, inproceedings}.

The remainder of this paper is structured as follows: Section~\ref{sec:related_works} provides a brief overview of the most relevant works in the field of QRC. In Section~\ref{sec:methodology}, we describe the methodology used in this work, including a description of the dataset and the QRC architecture. In Section~\ref{sec:results}, we present the results obtained from our experiments, encompassing both emulation on classical resources and \textit{Aquila}'s Quantum Processing Unit (QPU) simulations. Finally, in Section~\ref{sec:conclusions}, we draw conclusions and discuss future work.

\section{Related works} \label{sec:related_works}

Reservoir Computing (RC) is a computational framework designed to take advantage of the dynamic properties of Recurrent Neural Networks (RNNs) for efficient time series processing~\cite{10.5555/2968618.2968694, 2023GeoRL..5002649A} and pattern recognition~\cite{7311148, 7982291}. Unlike traditional RNNs, RC approaches keep the recurrent connections within a high-dimensional, nonlinear fixed "reservoir", reducing the complexity of training. This efficiency comes from the fact that only the output layer is trained, typically a simple classifier with relatively few parameters. As a result, RC models are not only lightweight but also easier to optimize. The reservoir itself functions as a dynamic memory, encoding input signals through rich internal states that capture temporal dependencies and complex patterns. This makes RC particularly effective and scalable for applications such as signal processing, forecasting, and chaotic time series prediction. This architecture has demonstrated robustness and adaptability in various domains, from neuroscience-inspired computing to hardware implementations in photonic and memristive systems~\cite{TANAKA2019100}.\\
QRC was initially proposed by Fujii and Nakajima as a resource-efficient framework for quantum machine learning, utilizing the natural evolution of quantum states as a high-dimensional computational reservoir with minimal training overhead~\cite{PhysRevApplied.8.024030}.
Some architectures have also been proposed for digital quantum computers. In particular, Yasuda et al. demonstrated a QRC approach based on repeated quantum measurements on superconducting qubits~\cite{Yasuda:2023vop}, while Kobayashi et al. proposed a feedback-driven QRC system to mitigate the destructive effects of quantum measurements~\cite{PRXQuantum.5.040325}.\\
However, few works explored the potential of analog simulators. Martínez-Peña et al. examined how phases such as thermalization and many-body localization affect the information processing capabilities of QRC~\cite{Mart_nez_Pe_a_2021}. In the field of neutral atom platforms, Bravo et al. implemented a quantum version of recurrent neural networks (qRNNs) using Rydberg atom arrays, which exhibit abilities to learn cognitive tasks such as multitasking, decision-making, and long-term memory retention~\cite{PRXQuantum.3.030325}.\\
In this context, Kornjača et al.\cite{kornjača2024largescalequantumreservoirlearning} demonstrated the large-scale applicability of QRC to neutral atom platforms with implementations using up to 108 qubits and benchmarking them on some well-known datasets. The authors introduced three encoding typologies: global pulse, local detuning, and position. The latter two showed better performances for the tasks considered and therefore will be implemented and tested in this work, as detailed in Section~\ref{sec:methodology}. However, it is still not clear which type of data and machine learning (ML) tasks are suitable for QRC methods, and real-world applicability is still to be assessed. This study aims to contribute to a better understanding of these aspects.
As mentioned in Section~\ref{sec:introduction}, we chose as a test bench a well-documented classification problem in finance, credit card default prediction, and identified a dataset that is frequently referenced in the literature~\cite{default_of_credit_card_clients_350}.
In the study by~\cite{YEH20092473}, for example, the authors compared the performance of six data mining methods-K-nearest neighbor classifier (KNN), logistic regression (LR), discriminant
analysis (DA), Na\"ive Bayes classifier (NB), artificial neural network (ANN) and classification trees (CT)-to predict cases of credit card default in Taiwan, using the same imbalanced dataset~\cite{default_of_credit_card_clients_350} used in this article.
A comparison of different machine learning approaches is also presented in~\cite{islam2018creditdefaultminingusing}, where K-Nearest Neighbors (KNN), Random Forest, Na\"ive Bayes (NB), Gradient Boosting and Extremely Random Trees (Extra Trees) are evaluated on the same dataset~\cite{default_of_credit_card_clients_350} using Accuracy, Recall, F-score, and Precision as metrics. 
Other studies~\cite{XIONG2013665} have focused on clustering techniques to discover patterns, and some other research~\cite{inproceedings} has explored online learning methods to update models, in real time, based on new data, with the aim of reducing the computational time, storage requirements, and processing overhead associated with retraining.

It should be noted that, among the various classification models discussed in the related work, we chose to apply to the reservoir-transformed features only those compatible with the QRC principle of simple training. For comparing our proposed methodology with classical nonlinear approaches, we included a Deep Neural Network (DNN) as a benchmark.

\section{Methodology and Algorithm design}\label{sec:methodology}
This section details the design and implementation of the hybrid quantum-classical pipeline developed for credit card default prediction. The proposed methodology integrates data preprocessing, quantum and classical reservoir computing, and a variety of classification strategies to address the challenges posed by imbalanced real-world datasets. In particular, we define a pipeline composed of three steps, each of which has been designed to maximize the extraction of informative features and to enable a fair comparison between quantum and classical approaches.\\
First, we provide a high-level view of the implemented pipeline.
\begin{itemize}
    \item \textbf{Preprocessing}: the dataset is preprocessed using standard methodologies in order to obtain a set of relevant features from a cleaned dataset;
    \item \textbf{Quantum Reservoir Computing}: this quantum layer takes the preprocessed features as input and maps them into a higher-dimensional space to better capture relationships among the original features.
    \item \textbf{Classification}: a few simple models, combined with resampling techniques, perform the binary classification task.
\end{itemize}

\begin{figure}[htbp]
    \centering
    \includegraphics[width=0.9\textwidth]{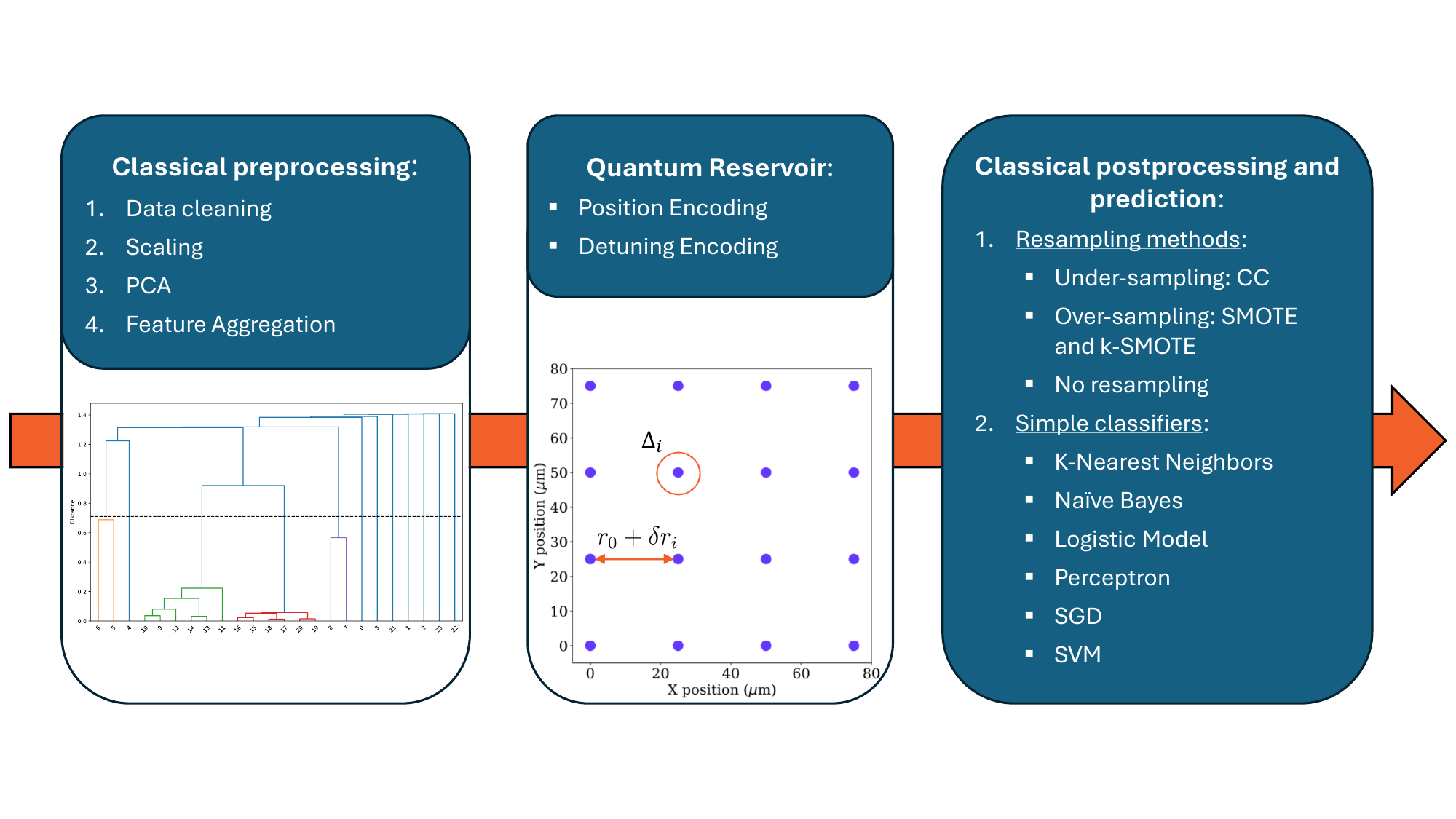}
    \caption{High-level view of the hybrid quantum-classical classification pipeline.}
    \label{fig:pipeline}
\end{figure}

A high-level diagram of the pipeline is shown in Fig.~\ref{fig:pipeline}, which also summarizes the options tested at each step.

Additional benchmarking is also performed using a classical approach, which considers two scenarios:
\begin{itemize}
    \item Classical Reservoir Computing (CRC): the whole pipeline remains unchanged, while the quantum reservoir computing layer is replaced by a classically-equivalent counterpart.
    \item Preprocessing only: the classifiers are applied directly to the features obtained after the preprocessing phase.
    This pipeline has been implemented for two purposes: to evaluate the contribution of the QRC module to the performance of the aforementioned simple classifiers and to assess the overall approach against a non-linear classical target (a DNN model). To better distinguish between these cases, we refer to the former, where the simple classifiers are applied directly to the preprocessed features, as fully-classical (FC).
\end{itemize}

In the following subsections, we describe the dataset, preprocessing procedures, reservoir computing layers, resampling techniques, and classification models adopted in this study. Further details can be found in the Appendix.

\subsection{Benchmark Dataset}
Credit card default prediction is a well-documented case study for which public datasets are readily available and have been used to benchmark machine learning techniques. For this specific use case, we use a public dataset~\cite{default_of_credit_card_clients_350} that contains information about credit card clients in Taiwan, covering the period April 2005 to September 2005.\\
The original dataset consists of 30,000 samples, each with 23 features and a binary label indicating whether a credit card was defaulted in the following month. Each sample corresponds to a credit card, and the 23 features encompass various aspects: credit data (amount of credit extended in NT dollars), demographic factors (sex, education, marital status, and age), payment history (six values representing repayment status from September 2005 to April 2005, where -1 indicates timely payments and values from 1 to 9 indicate delays ranging from one to nine months), bill statement history (six values detailing the amounts of the bill statements from September to April) and previous payment history (six values reflecting the amounts paid in the previous months, from September to April). More details on the dataset can be found in Appendix~\ref{subsec:cleaning_preprocessing}.

\subsection{Preprocessing}

Our classification pipeline begins with a preprocessing stage designed to provide the reservoir with the most informative features possible. We started by analyzing our dataset and assessing its overall integrity. Therefore, the initial step in our preprocessing is \emph{data cleaning}, removing negligible samples whose features appear to have undocumented values. The details of the process are reported in the Appendix~\ref{subsec:cleaning_preprocessing}.

The cleaned dataset consists of 29601 samples, and will be referred to as the \texttt{CARDS\_30000} dataset for brevity. Due to limited quantum resources, the entire dataset was not executed on \textit{Aquila}. Instead, two additional nested subsets were run on the quantum hardware: \texttt{CARDS\_2500}, containing 2571 samples, and \texttt{CARDS\_1000}, containing 1000 samples. Both subsets were stratified according to class labels to preserve class distribution. These datasets are nested as follows:
\[
\text{\texttt{CARDS\_1000}} \subseteq \text{\texttt{CARDS\_2500}} \subseteq \text{\texttt{CARDS\_30000}}
\]
 However, this restriction allows us to evaluate the effectiveness of the QRC method in scenarios with limited data availability, an expected advantage of reservoir computing, which typically requires fewer data for training~\cite{gauthier2021next}.

The dataset contains two categorical variables \textit{SEX} and \textit{MARRIAGE} that we addressed by converting them into boolean features, as they do not have a natural ordering. This transformation increases the total number of features from 23 to 24.
Then, we apply a series of scaling and PCA (Principal Component Analysis) steps, with the ultimate goal of supporting feature extraction through the correlation circle method~\cite{james2013introduction}. 
While this may appear to contradict the QRC step, which increases dimensionality, it actually serves a dual purpose. First, it eliminates non-informative features and aggregates those carrying highly correlated information. Second, it allows for a more efficient encoding of samples on the QPU, whether through detuning or position encoding. In fact, having fewer features enables multiple replications of the same sample within the QPU register, thereby reducing the quantum resources required (more details are provided in Section~\ref{sec:results}).

\begin{figure}[htbp]
    \centering
    \includegraphics[width=0.9\textwidth]{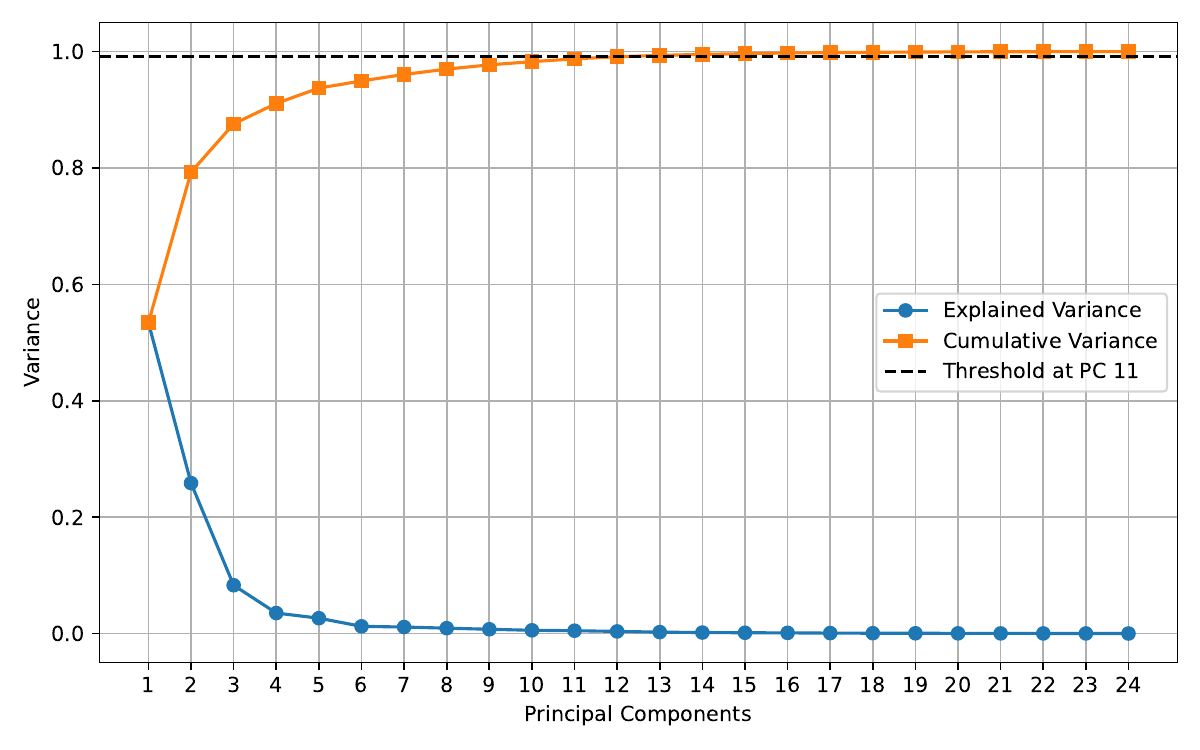}
    \caption{PCA explained variance, computed on the \texttt{CARDS\_30000} dataset.}
    \label{fig:variance}
\end{figure}

\begin{figure}[htbp]
    \centering
    \includegraphics[width=0.9\textwidth]{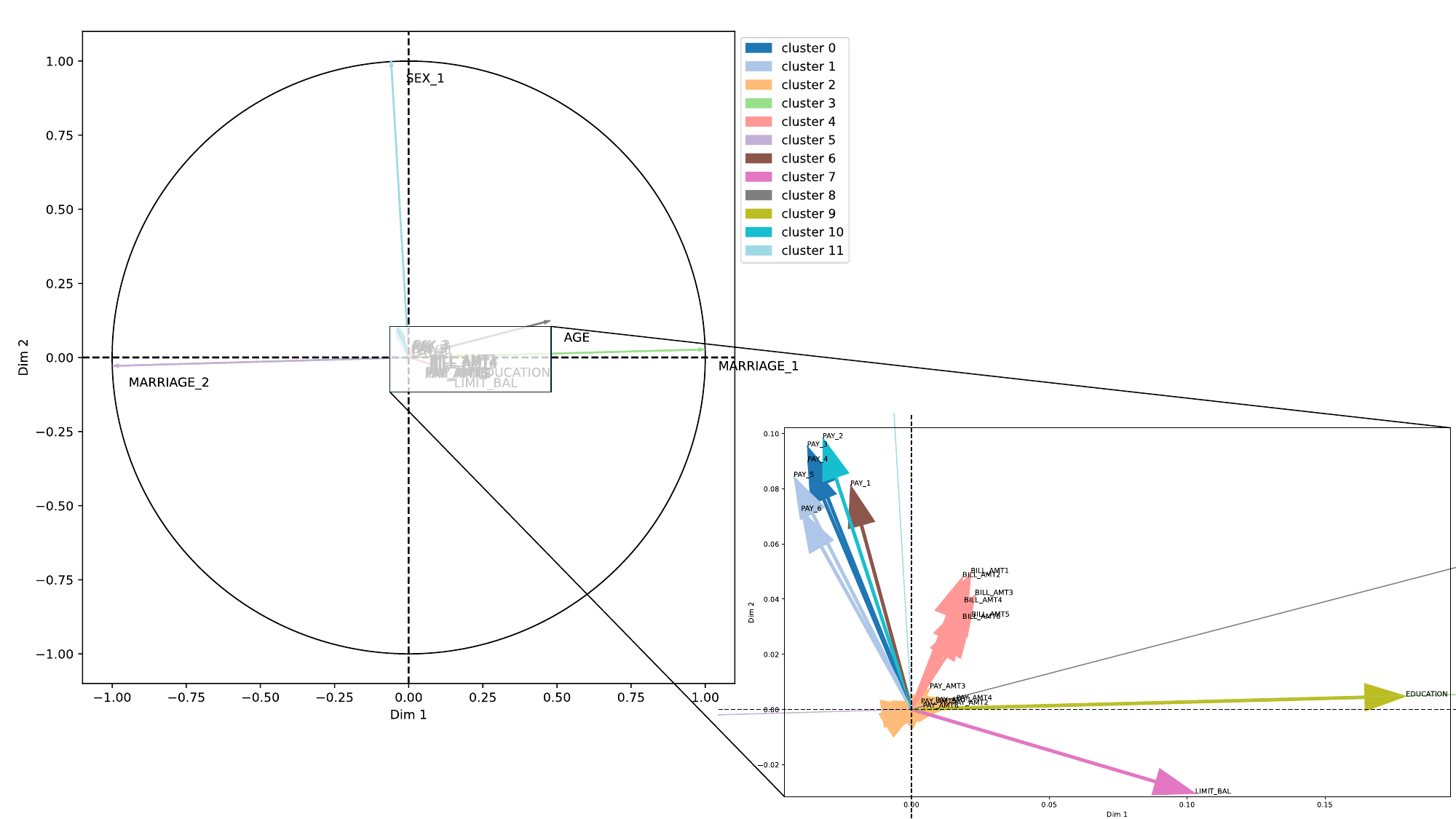}
    \caption{Features aggregation on the \texttt{CARDS\_30000} dataset.}
    \label{fig:corr_circle}
\end{figure}

Fig.~\ref{fig:variance} illustrates the PCA process, which results in the selection of 11 principal components, which approximately explain 99\%  of the total variance in the dataset. Based on the principal component scores associated with each feature, we then aggregate the 24 features into 12 groups using hierarchical clustering in the principal component space. Fig.~\ref{fig:corr_circle} displays each feature vector projected into the 2D space defined by the first two principal components, where the features belonging to the same cluster are indicated by the same color.

So, at the end of the preprocessing step, we obtain a reduced number of features, that is 12 in our case.

The subsequent reservoir computing step then increases the dimensionality of the data with the specific objective of capturing non-linear dependencies among features-interactions that cannot be identified by the linear methods applied during the preprocessing stage.

\subsection{Reservoir Computing}

\subsubsection{Quantum Reservoir Computing}\label{subsec:QRC}

As detailed in Section~\ref{sec:related_works}, a variety of Quantum Reservoir Computing approaches have been proposed in recent years. In terms of scale,~\cite{kornjača2024largescalequantumreservoirlearning} have demonstrated execution on neutral atom platforms in the hundreds-qubit range with a possible advantage over classical ML methods in terms of scalability. For this reason, a similar approach has been applied in this classification use-case. To execute a QRC routine, it is required to encode the classical data obtained after the feature aggregation process. For the considered quantum platform \textit{Aquila}, this can be achieved using two different methodologies.
In both setups, the qubits (atoms) representing a sample from the dataset are logically arranged along a one-dimensional array, ensuring that interactions, used to encode feature relationships, are primarily limited to nearest neighbors.\\
The encoding methodologies, using the notation of Equation~\eqref{hamiltonian_eq}, are described as follows:
\begin{itemize}
    \item \textit{position encoding}, this approach manipulates the Rydberg interactions strength between neighboring atoms by varying the inter-atomic distances depending on the data features. In particular, for a given sample, the separation between the $i^{th}$ and $(i+1)^{th}$ atoms is given by $r_{i,i+1} = r_0 + \delta r_i \rightarrow r_0 (1+\lambda f^i)$. Here, $\delta r_i$ is the displacement of the $i^{th}$ and the $(i+1)^{th}$ atoms with respect to the fixed distance $r_0$, $f^i$ is the $i^{th}$ feature of the considered sample, and finally $\lambda$ is a displacement scaling factor that influences $\delta r_i / r_0$. Consequently, encoding an $n$-dimensional sample requires an array of $n+1$ qubits (atoms). Within this scheme, a uniform global detuning drive is applied, i.e., $\Delta_i(t) = \Delta_{global}(t), \ \forall i$, with no site-dependent local detuning;
    \item \textit{detuning encoding}, this strategy directly maps the $i$-th feature $f^i$ to the amplitude of the local detuning drive: $\Delta_{i}(t) \rightarrow \Delta_{global}(t) + \Delta_{l} f^i$. Here, $\Delta_{global}$ is a fixed global detuning drive, and $\Delta_{l}$ is the scaling coefficient of the local detuning pulse. For this method, the number of atoms (qubits) in the array is equal to the number of features, and the inter-atomic distance remains fixed: $r_{i,i+1} = r_0, \ \forall i$.
\end{itemize} 

Both approaches are visualized in the second block of Fig.~\ref{fig:pipeline}.\\
In order to construct the output quantum feature space, the expectation values of $Z_i$ and $Z_i Z_j$ operators for each pair of atoms must be computed for a set of timesteps. To this end, an initial state $\ket{\psi_{t=0}} = \ket{0\ldots0}$ is prepared. Then, the system evolves following Eq.~\eqref{hamiltonian_eq} and $\bra{\psi}Z_i\ket{\psi}$ and $\bra{\psi}Z_i Z_j\ket{\psi}$ are computed at each timestep.
Hence, the whole set represents the QRC output for one sample.

So, in our case, given \( n = 12 \) features resulting from the preprocessing step, the detuning encoding scheme produces 
\(
n + \binom{n}{2} = 12 + \binom{12}{2} = 12 + 66 = 78
\)
features per time step, accounting for both single-feature $\bra{\psi}Z_i\ket{\psi}$ and pairwise (coupled) correlations $\bra{\psi}Z_i Z_j\ket{\psi}$. Considering 5 time steps, this results in a total of 
\(
78 \times 5 = 390
\)
output features.

In the case of position encoding, one additional qubit is required to represent the features. Thus, the number of features per time step becomes
\(
(n + 1) + \binom{n + 1}{2} = 13 + \binom{13}{2} = 13 + 78 = 91,
\)
yielding a final dimensionality of
\(
91 \times 5 = 455
\)
output features. 

The reservoir output features are then used to feed the classifiers to obtain the final label for the data sample.

\begin{figure}[htbp]
    \centering
    \begin{subfigure}[b]{0.45\textwidth}
        \centering
        \includegraphics[width=\textwidth]{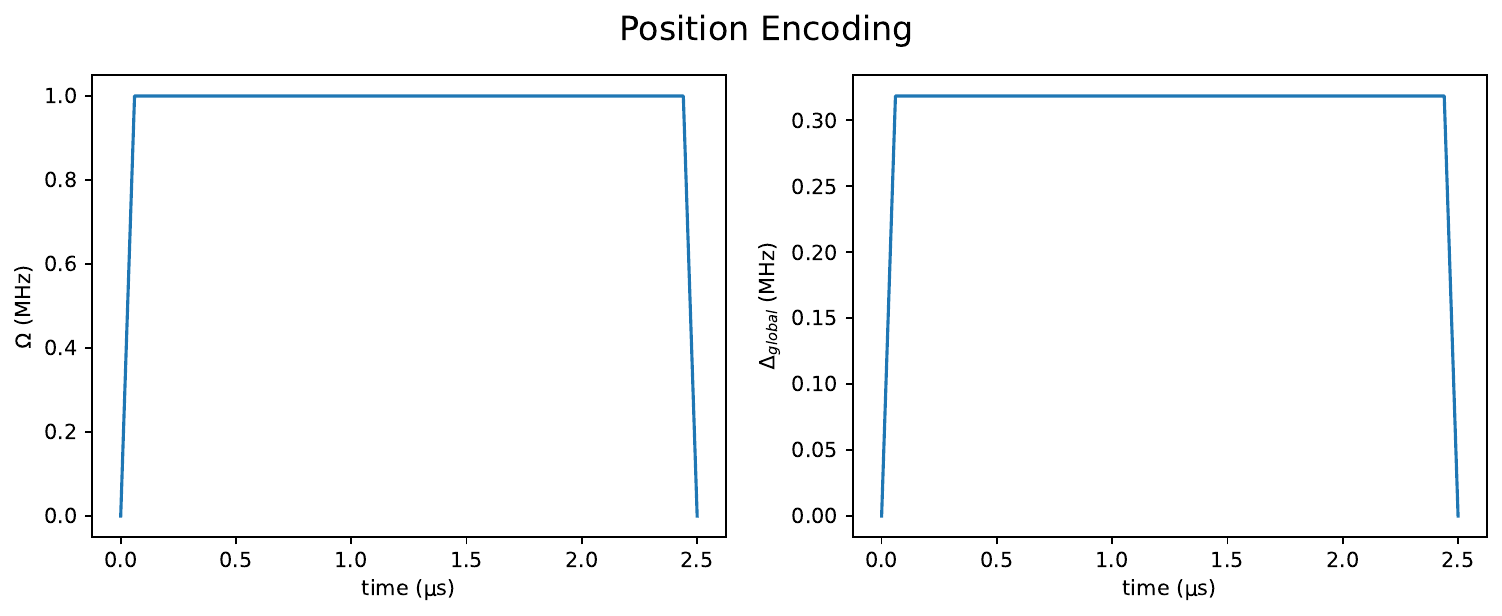}
        \caption{}
        \label{fig:waveform_position_encoding}
    \end{subfigure}
    \begin{subfigure}[b]{0.45\textwidth}
        \centering
        \includegraphics[width=\textwidth]{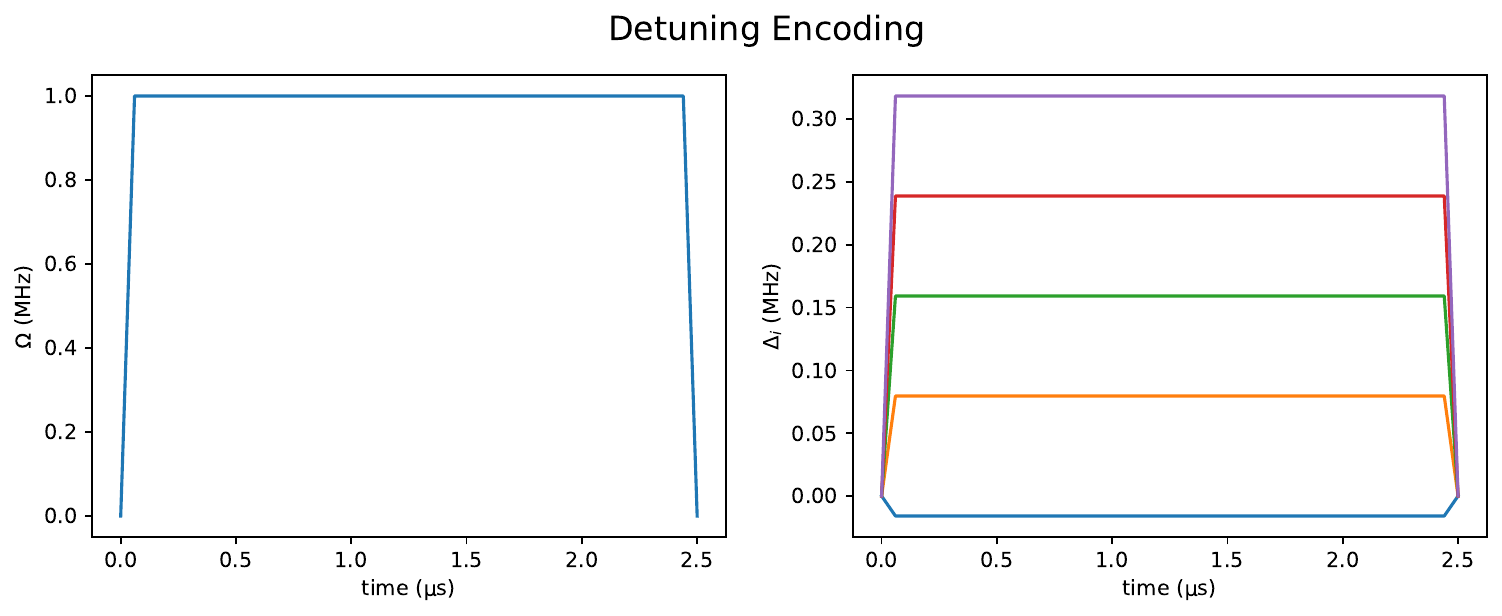}
        \caption{}
        \label{fig:waveform_detuning_encoding}
    \end{subfigure}
    \caption{Example pulse sequence for the (a) position encoding and (b) detuning encoding. In the latter case, we considered a short 5 qubit array for the sake of readability. The relative local detuning waveforms are plotted using different colors.}
    \label{fig:det_pulse_emu}
\end{figure}

To define pulse sequences, the profiles of both $\Omega(t)$ and $\Delta_i(t)$ must be defined (see~\eqref{hamiltonian_eq}). The same waveform profile has been used for both parameters. In particular, we implemented a steep $0.05 \ \mu s$ ramp up and ramp down to a maximum value, as reported in Fig.~\ref{fig:det_pulse_emu}. In order for the Hamiltonian simulation to be within the typical coherence time of such systems ($\sim4 \ \mu s$) and to minimize the impact of noise on the result, we decided to use 5 timesteps of $0.5 \ \mu s$ each.

\subsubsection{Classical Reservoir Computing}
In order to assess the potential of the reservoir computing methodology per se, we also tested with a classical equivalent of the QRC. Specifically, the Classical Reservoir Computing (CRC) benchmark is a \textit{dequantized} version of the QRC detailed in Section~\ref{subsec:QRC}, obtained by mapping the qubits to classical unit vectors $\hat{\mathcal{S}}$, as defined in~\cite{kornjača2024largescalequantumreservoirlearning}.
In particular, from the quantum system described by~\eqref{hamiltonian_eq}, we retrieve the equivalent classical system defined by: 
\begin{equation}\label{crc_eq}
    \frac{\partial\hat{\mathcal{H}}}{\partial\hat{\mathcal{S}_i}} = \frac{\Omega(t)}{2} \hat{x} +\left[ -\frac{\Delta_i(t)}{2} + \sum_{i \neq j} \frac{V_{ij}}{4} \left( 1 + \hat{\mathcal{S}}_j^{(z)} \right) \right] \hat{z}.
\end{equation}

The Hamiltonian evolution of the $N$ classical spins can be emulated through numerical solvers ($3N$ equations). Using the same set of parameters defined in the QRC procedure, we can compute the spin projection on the \textit{z-axis} and their products at each timestep.

The use of the module in the pipeline is equivalent to the quantum version: the preprocessed features are fed into the classical reservoir module, whose output features are then used as input for the classifiers.

\subsection{Classification}

The final stage of our pipeline focuses on the binary classification of credit card defaults. This stage comprises two key components: \textit{1)} resampling techniques to address class imbalance, and \textit{2)} classification models to predict default status.

\subsubsection{Resampling Methods}

The datasets used in this study are significantly imbalanced, with the ratio of defaulters (class 1) to non-defaulters (class 0) being approximately $0.22$. This imbalance is consistent across all three datasets \texttt{CARDS\_30000}, \texttt{CARDS\_2500}, and \texttt{CARDS\_1000}, in accordance with the stratified approach used to create the smaller datasets.

To mitigate the effects of this imbalance, we introduced a resampling phase prior to model training. Specifically, we evaluated both oversampling and undersampling techniques:

\begin{itemize}
    \item \textbf{Oversampling} artificially increases the number of minority class samples (defaulters) by generating synthetic instances based on existing class 1 samples. This helps the model to better learn the characteristics of the minority class.
    \item \textbf{Undersampling} reduces the number of majority class samples (non-defaulters) by selecting only the most informative examples, thus balancing the dataset by bringing the majority class down to the size of the minority class.
\end{itemize}

All resampling methods were applied exclusively to the training set to avoid data leakage. The validation and test sets were left unchanged to ensure an unbiased evaluation of model performance on real, unmodified data.

We experimented with the following resampling strategies:
\begin{itemize}
    \item \textit{Synthetic Minority Over-sampling Technique (SMOTE)} and \textit{K-Means SMOTE (K-SMOTE)}~\cite{chawla2002smote} for oversampling,
    \item \textit{Cluster Centroids (CC)}~\cite{lin2017clustering} for undersampling,
    \item A baseline case with no resampling (labeled as \textit{None}), allowing the models to handle the imbalance inherently.
\end{itemize}

Further details of the implementation and a deeper explanation of these techniques are provided in Appendix~\ref{apx:resampling}.

\subsubsection{Classifiers}

In this study, a variety of classification algorithms were used to evaluate their effectiveness in handling the underlying data distribution. These models were selected based on their ease of training, alignment with the principles of reservoir computing, and broad support within the \texttt{scikit-learn} Python library~\cite{scikit-learn}. 

Hyperparameter tuning played a central role in optimizing model performance. For models requiring parameter calibration, a range of candidate values was explored using a grid search approach. The selection criterion was the F1-score on the validation set, a metric well-suited for imbalanced classification tasks, as it balances precision and recall. In contrast, accuracy was deliberately avoided because of its tendency to provide misleadingly high values in the presence of class imbalance. Where applicable, class weights were incorporated to further mitigate the bias toward the majority class.

Table~\ref{tab:hyperparams} summarizes the classification models used in the study, the specific hyperparameters that were tuned, the hyperparameters values tested, and whether class weighting (CW) was applied. More details about the classifiers can be found in the Appendix~\ref{apx:classifiers}.

\begin{table}[h!]
\centering
\caption{Hyperparameter configurations}
\begin{tabular}{|l|p{2.2cm}|p{2.5cm}|c|}
\hline
\textbf{Model} & \textbf{Hyperparameter} & \textbf{Values Tested} & \textbf{CW} \\
\hline
KNN & $k$ & $\{2, 3, \dots, 15\}$ & No \\
\hline
SGD & Epochs & $\{10, 20, \dots, 100\}$ & Yes \\
\hline
Naïve Bayes & None & -- & No \\
\hline
Log. Model & $C$ & $\{10^{-3}, \dots, 10^2\}$ & Yes \\
\hline
Perceptron & $\eta$ & $\{10^{-1}, 10^{-2}, 10^{-3}\}$ & Yes \\
\hline
SVM & $C$ & $\{10^{-1}, 10^0, 10^1, 10^2\}$ & Yes \\
\hline
\end{tabular}
\label{tab:hyperparams}
\end{table}

\subsection{Deep Learning model}\label{subsubsec:dnn}

The Deep Neural Network (DNN) model used in this work is designed for classification tasks and is implemented using \texttt{PyTorch}~\cite{imambi2021pytorch}. The model architecture consists of six fully-connected layers with dropout regularization to prevent overfitting. The input layer size corresponds to the dimensionality of the feature space, followed by hidden layers of varying sizes (64, 128, 256, 128, 64 neurons) and \textit{ReLU} activation functions. The output layer consists of two neurons for binary classification.

The DNN is applied directly to the preprocessed features, removing any reservoir layer from the pipeline, to assess the performance of the proposed methodology with respect to a deep learning approach.
The training process involves optimizing the cross-entropy loss function using the \textit{Adam} optimizer. The dataset is split into training, validation, and test sets, where the training data is used for model fitting, and the validation data guides hyperparameter selection. The training loop iterates over 1000 epochs, computing loss and updating weights using backpropagation. An early stopping mechanism with a patience of 30 epochs is implemented to prevent overfitting.\\
The model undergoes hyperparameter tuning for two key parameters:
\begin{itemize}
\item \textbf{Learning rate:} Selected from ${0.1, 0.01, 0.001}$.
\item \textbf{Batch size:} Chosen from ${64, 128, 256}$.
\end{itemize}
A grid search over these parameters identifies the best-performing configuration based on the validation set F1-score. The best model is then evaluated on the test set to obtain the final classification performance.

\section{Results}\label{sec:results}

In our work, we began by performing extensive emulations of the quantum system using classical computational resources on the \texttt{CARDS\_30000} dataset. This preliminary phase allowed us to evaluate the performance of our method in a controlled, noiseless environment, providing a clear baseline for comparison and aiding in the identification of suitable parameters for our approach. After validating the method through emulation, we proceeded to test it on \textit{Aquila}'s QPU, using \texttt{CARDS\_2500}, \texttt{CARDS\_1000} datasets.

\subsection{Emulation results}\label{sec:emulation}

In this section, we present the results of the emulation campaign conducted to assess the classification scores and training times of the proposed methodolodgy, as well as the definition of the proper parameters for detuning and position encoding. 
Since both encoding strategies are expected to exhibit similar behavior under emulation, performance evaluations on the \texttt{CARDS\_30000} dataset were conducted exclusively using the detuning encoding. However, to ensure a fair comparison on the smaller \texttt{CARDS\_1000} dataset, which was tested on real hardware to assess the effectiveness of both encoding schemes, position encoding was also subjected to emulation.
The noiseless classical emulation of the quantum system has been performed using the Julia version of the \textit{Bloqade}~\cite{bloqadejl} library.

In all emulation and encoding setups, we set $r_0 = 10 \ \mu m$ and a maximum Rabi frequency at the end of the ramp up of $\Omega_{max} = 2\pi \ rad/ \mu s$, based on the typical values reported in~\cite{kornjača2024largescalequantumreservoirlearning}.\\
In addition to the parameters shared between the two approaches, specific optimizations were performed for each encoding scheme.
For the \textbf{position encoding}, both the scale parameter $\lambda$ and the global drive $\Delta_{global}$ were tuned. The optimal configuration $\lambda = 1.0$ and $\Delta_{global} = 2.0\ \text{rad}/\mu s$ was selected as it provides a well-balanced trade-off across key performance metrics and aligns effectively with the capabilities of \textit{Aquila}'s QPU.
On the other hand, when looking at the \textbf{detuning encoding}, varying the global detuning value $\Delta_{global} \in [0, 9]\ rad/\mu s$ does not significantly affect the classification performances. In this case, we set $\Delta_{global} = \pi \ rad/\mu s$. Fig.~\ref{fig:det_pulse_emu} reports the pulse sequence for both types of encoding.

Since previous work suggests that detuning and position encoding approaches yield comparable performance in a noise-free emulation setting~\cite{kornjača2024largescalequantumreservoirlearning}, and given the high computational cost of classically emulating quantum systems, we focused our efforts regarding \texttt{CARDS\_30000} dataset's results, within the \emph{detuning} encoding framework.

\begin{figure}[h!]
    \centering
    \includegraphics[width=0.9\textwidth]{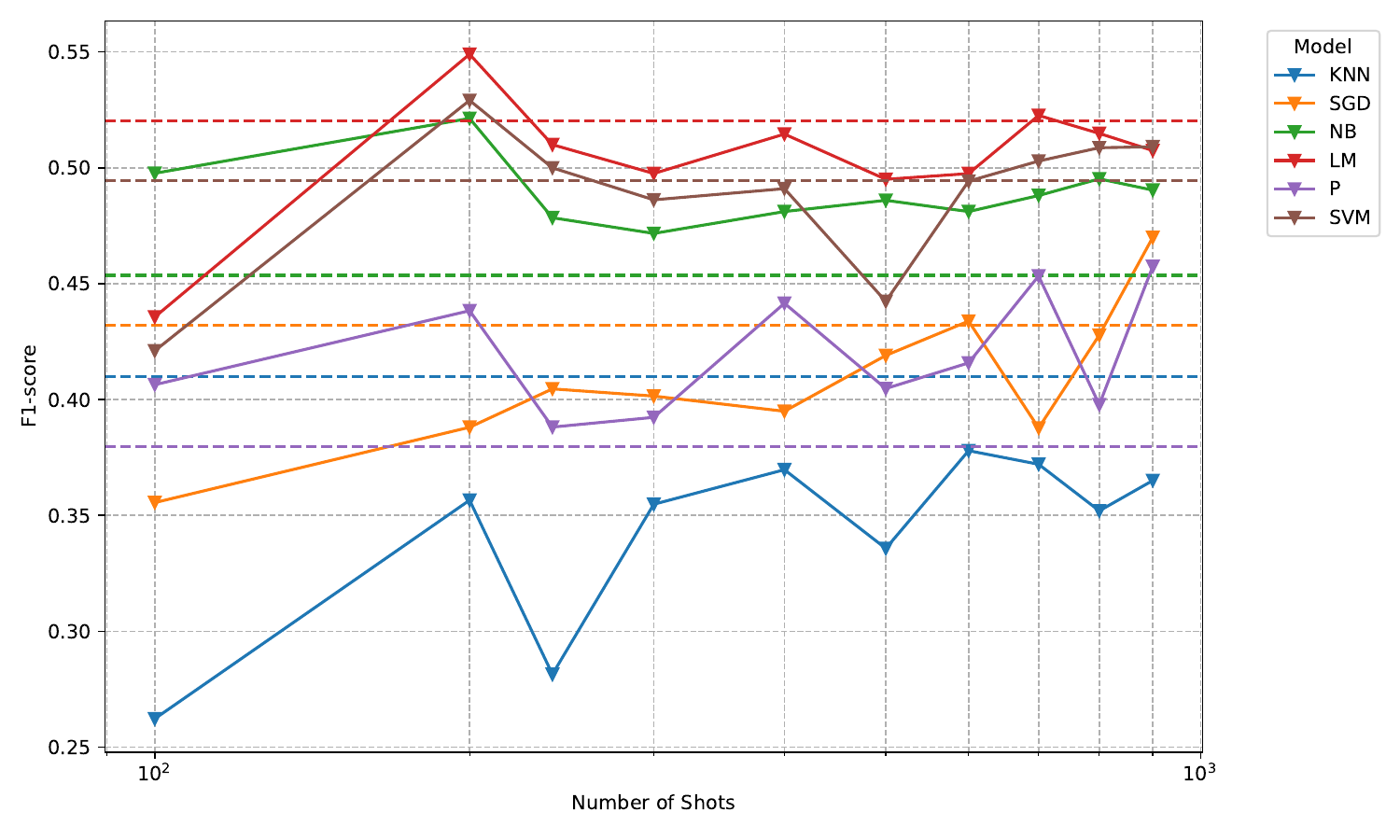}
    \caption{F1-score as a function on the number of Hamiltonian simulations. Dotted lines refer to classification scores on the noiseless statevector emulation of the QRC procedure.}
    \label{fig:shot_opt_F1}
\end{figure}

Concerning the calculation of QRC features $\bra{\psi}Z_i\ket{\psi}$ and $\bra{\psi}Z_i Z_j\ket{\psi}$ have been calculated both using statevectors (sv in the following figures) and shot-based measurements.
The latter has been used to estimate the appropriate number of shots to minimize the loss of classification performance due to statistical noise.\\
Fig.~\ref{fig:shot_opt_F1} reports the F1-score as a function of the number of measurements. The performance seems to reach a plateau when performing a few hundred emulated Hamiltonian evolutions.

\begin{figure*}[h!]
    \centering
    \includegraphics[width=0.95\textwidth]{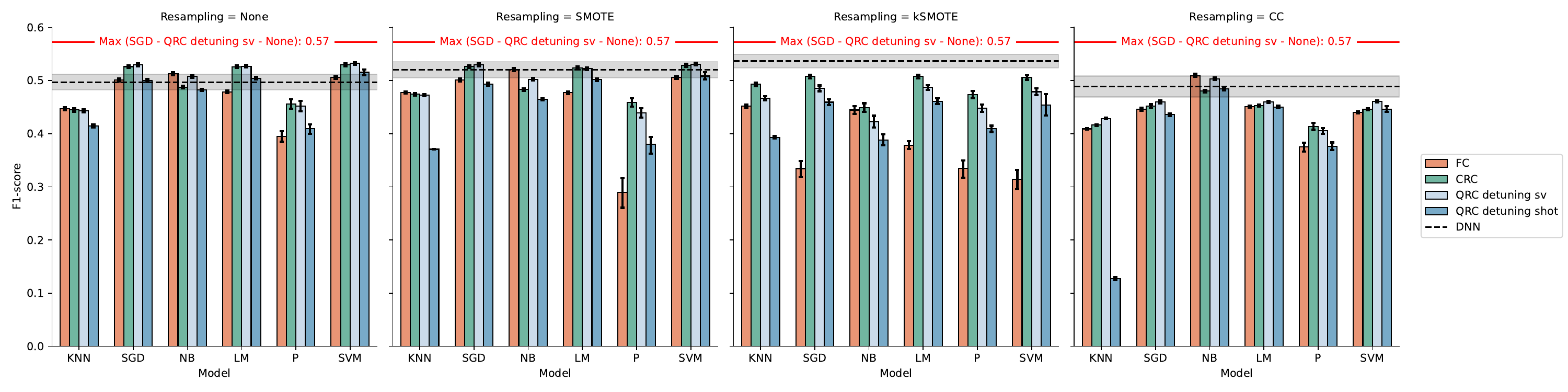}
    \caption{Benchmark of the classification results on \texttt{CARDS\_30000}, which compares emulated QRC (detuning encoding), CRC, and FC methods against the DNN. Dark shaded area shows the standard deviation for the DNN, while the red line indicates the maximum performance reached by any of the considered classifier.}
    \label{fig:full_dataset_results}
\end{figure*}

According to the methodology described in Section~\ref{sec:methodology}, we performed emulations to preliminarily assess the goodness of the reservoir methods in terms of both classification performance and training times.
As shown in Fig.~\ref{fig:full_dataset_results}, the resulting F1-score for all considered reservoir models is compatible with the performance achieved by DNN, particularly for statevector emulations.

Table~\ref{tab:timing} instead reports the training times of the DNN and of the other classifiers when using the QRC output feature vectors.

\begin{table*}[ht!]
\centering
\caption{Comparison of the training times (in seconds) of the considered classifiers on \texttt{CARDS\_30000} for different resampling methods. The DNN is trained on the original 12 features, while the other classifiers use the QRC output features (detuning encoding). As expected, the training time increases when applying oversampling methods and a decreases for the undersampling one.}

\label{tab:timing}
\begin{tabular}{|l|c|c|c|c|}
\hline
\textbf{Classifier} & \textbf{None} & \textbf{SMOTE} & \textbf{kSMOTE} & \textbf{CC} \\ \hline
KNN                 & $0.12 \pm 0.01$ & $0.23 \pm 0.01$ & $0.24 \pm 0.11$ & $0.10 \pm 0.01$ \\ \hline
SGD                 & $2.86 \pm 0.34$ & $3.43 \pm 0.40$ & $2.82 \pm 0.31$ & $1.22 \pm 0.15$ \\ \hline
Na\"ive Bayes         & $0.11 \pm 0.01$ & $0.23 \pm 0.01$ & $0.20 \pm 0.01$ & $0.10 \pm 0.01$ \\ \hline
Logistic Model & $0.14 \pm 0.01$ & $0.31 \pm 0.02$ & $0.30 \pm 0.10$ & $0.08 \pm 0.01$ \\ \hline
Perceptron          & $0.18 \pm 0.02$ & $0.28 \pm 0.02$ & $0.27 \pm 0.01$ & $0.09 \pm 0.01$ \\ \hline
SVM                 & $1515.93 \pm 397.65$ & $4359.86 \pm 1076.96$ & $3976.44 \pm 1077.47$ & $150.71 \pm 83.96$ \\ \hline
\textbf{DNN}        & $26.52 \pm 0.26$ & $43.32 \pm 2.77$ & $42.97 \pm 1.77$ & $13.62 \pm 0.62$ \\ \hline
\end{tabular}
\end{table*}

With the exception of the SVM, the reservoir-enhanced classification shows faster training than the implemented DNN. In particular, the training time is one order of magnitude shorter with respect to the DNN for the SGD, and two orders of magnitude for the KNN, Na\"ive Bayes, Logistic Model and Perceptron.
Concerning SVM, training time is prohibitive with respect to other classifiers, because its training phase is highly affected by the dimensionality of the feature space. So, despite employing a linear kernel, the SVM does not conform to the RC principle of fast and efficient training.
The reduction of training times with respect to the DNN for all the other classifiers and the alignment of the F1-score motivate further investigation of the QRC approach using a real neutral atom quantum simulator.
The classification performance of the emulated data for the \texttt{CARDS\_2500} and \texttt{CARDS\_1000} datasets have been compared to real data on \textit{Aquila}'s QPU. Due to limitations in available quantum resources, tests on the QPU were conducted only on these smaller datasets. The detailed results are reported in Section~\ref{sec:exp_results}, however, also in those cases, the classification performance of the emulated statevector QRC procedures is comparable to DNN for both datasets, in line with the results obtained using the \texttt{CARDS\_30000} dataset.\\
In particular, in FC setups, reducing the number of dataset entries improves the performance of the NB classifier with respect to the DNN when CC resampling is used. This can be attributed to several factors: the CC representation reduces dataset size and noise, yielding a compact and discriminative feature space; the resulting features tend to exhibit low correlation, which, while not ensuring full conditional independence, supports the assumptions of Na\"ive Bayes; furthermore, the absence of hyperparameter tuning in Na\"ive Bayes allows it to train on the entire dataset without requiring a validation split, leading to a larger training set with respect to the other classifiers that are more exposed to overfitting.

\subsection{Experimental results}\label{sec:exp_results}

\begin{figure}[htbp]
    \centering
    \begin{subfigure}[b]{0.2\textwidth}
        \centering
        \includegraphics[width=\textwidth]{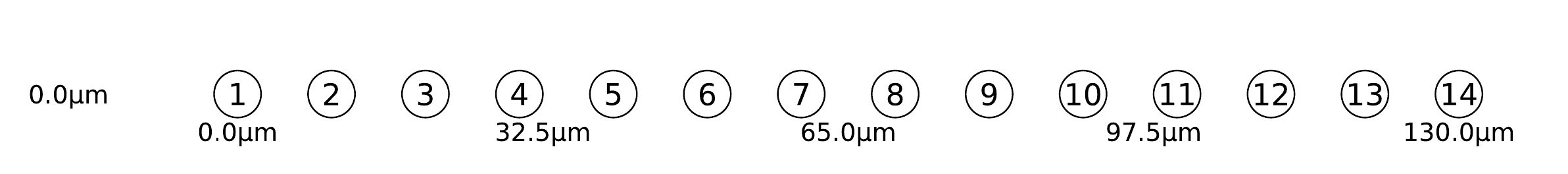}
        \caption{}
        \label{fig:single_hook_encoding_aquila}
    \end{subfigure}
    \begin{subfigure}[b]{0.2\textwidth}
        \centering
        \includegraphics[width=\textwidth]{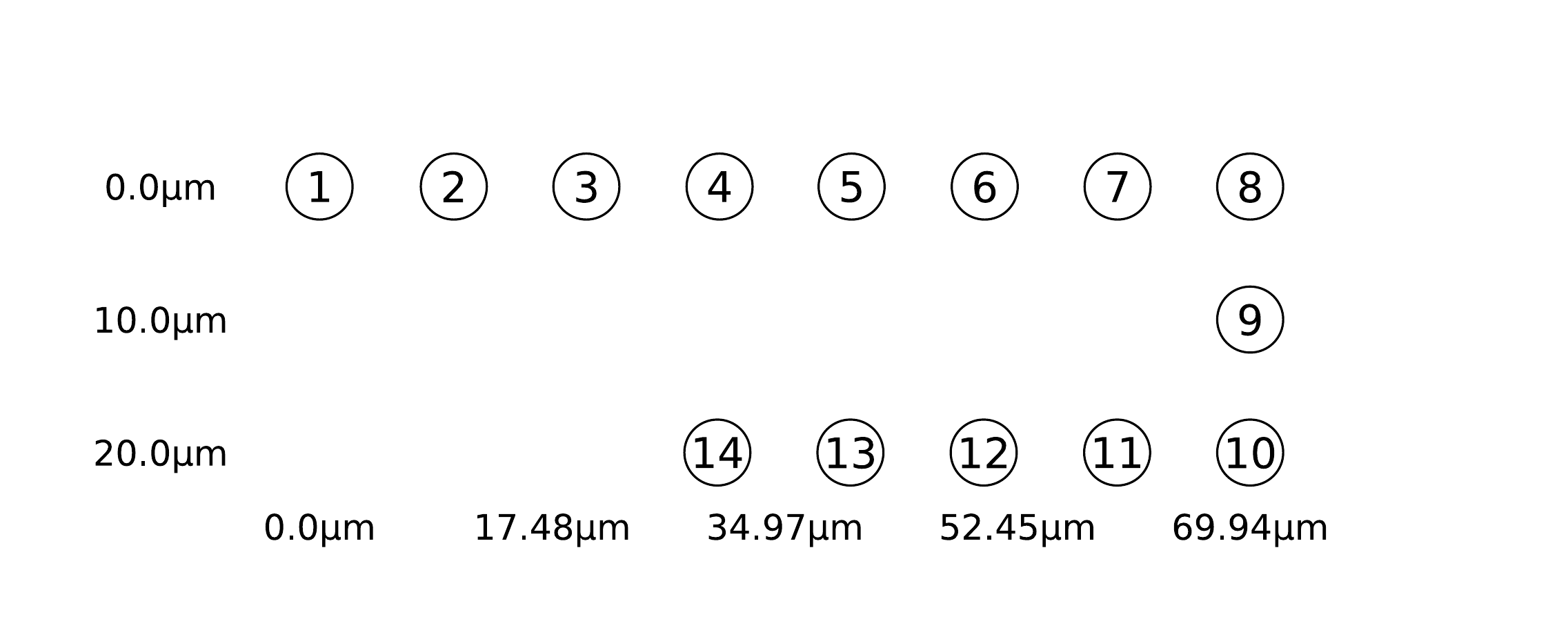}
        \caption{}
        \label{fig:single_column_encoding_aquila}
    \end{subfigure}
    \hspace{0.05\textwidth}
    \begin{subfigure}[b]{0.2\textwidth}
        \centering
        \includegraphics[width=\textwidth]{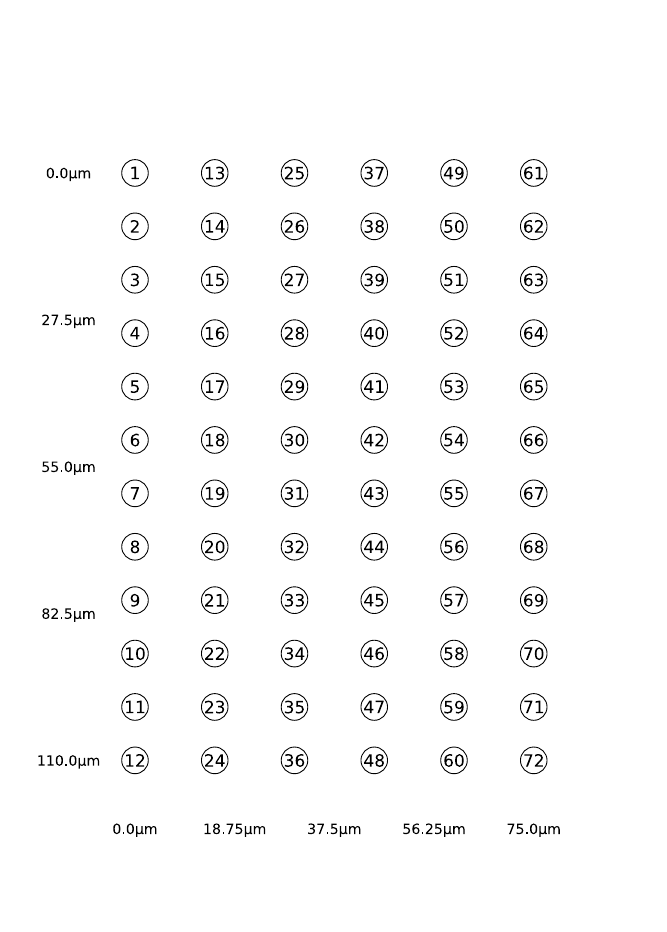}
        \caption{}
        \label{fig:column_encoding_aquila}
    \end{subfigure}
    \hspace{0.05\textwidth}
    \begin{subfigure}[b]{0.2\textwidth}
        \centering
        \includegraphics[width=\textwidth]{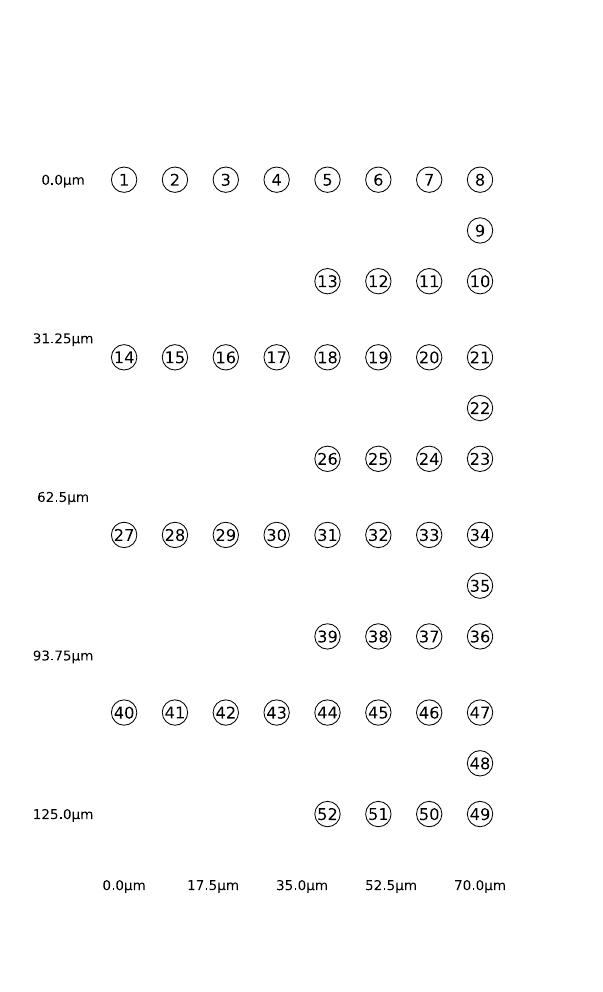}
        \caption{}
        \label{fig:hook_encoding_aquila}
    \end{subfigure}
    \caption{Comparison of different register encodings: (a) single column embedding (rotated by 90°), (b) serpentine-like embedding , (c) replicated column embedding (6 replicas) and (d) replicated serpentine embedding (4 replicas).}
    \label{fig:combined_encodings}
\end{figure}

Following extensive emulations, which were essential for defining all parameters and establishing a noiseless benchmark, we proceeded to experiment on a real neutral atom simulator. Due to its high availability, we employed the \textit{Aquila} platform from \textit{QuEra}. Notably, \textit{Aquila} is nowadays the only publicly accessible platform that supports local detuning modulation. 
A 2D register of size $75 \ \mu m \ \times \ 125 \ \mu m$ is provided to arrange the atoms within the machine. The vertical spacing and the Euclidean distance between the atoms should be at least $4 \ \mu m$, while there is no constraint on the horizontal spacing~\cite{wurtz2023aquilaqueras256qubitneutralatom}. Unlike the classical emulation of the quantum system, where atoms are assumed to be aligned in one dimension, the same qubit array must now be embedded within a two-dimensional register. In the considered case, if the array's length is smaller than the register's height, a straightforward column-wise embedding is applied. Otherwise, the array is mapped into a serpentine-like lattice.
Furthermore, depending on the register size and the length of the atom array, multiple replicas can be placed within the same register to improve the sampling rate. Examples of both column and serpentine embeddings, for single and multiple replicas, are shown in Fig.~\ref{fig:combined_encodings}.
For position encoding the atom array should be replicated by row rather than by column, if the array does not fit a single column. This is because, with a serpentine position embedding replicated by column, the vertical spacing between atoms is not guaranteed to meet the $4 \ \mu m$ requirement.

In addition, we decided to separate the replicas by $15 \ \mu m$ (more than the Rydberg radius and $r_0$) apart from each other to suppress cross-replica interactions.
In our case, both position and detuning encoding does not exceed the register's maximum height, so a column embedding strategy with $N_{repl} = 6$ replicas of the atom array has been implemented.\\
Another relevant difference from QRC emulation is related to the need to perform a quantum simulation for each timestep, restarting the QRC from $\ket{\psi_{t=0}}$, due to the inability to implement non-destructive measurements in the physical neutral atom platform. This translates into longer pulses with subsequent timesteps (e.g., $0.5 \ \mu s$ duration for the first timestep, $1 \ \mu s$ for the second one, etc.).

\begin{figure}[htbp]
    \centering
    \begin{subfigure}[b]{0.25\textwidth}
        \centering
        \includegraphics[width=\textwidth]{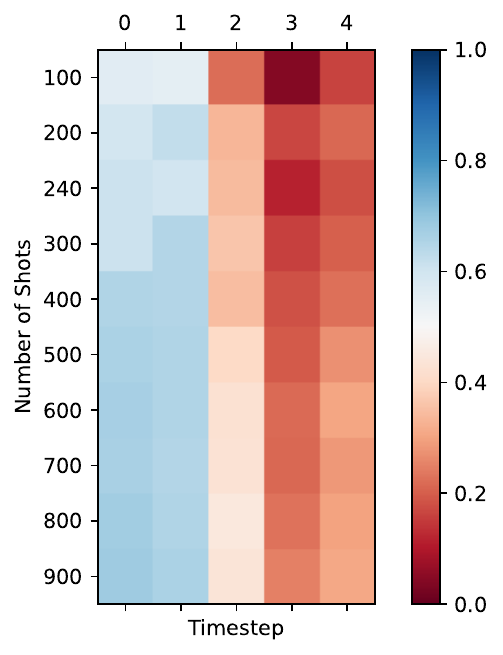}
        \caption{}
        \label{fig:aquila_correlation}
    \end{subfigure}
    \hfill
    \hspace{0.02\textwidth}
    \begin{subfigure}[b]{0.2\textwidth}
        \centering
        \includegraphics[width=\textwidth]{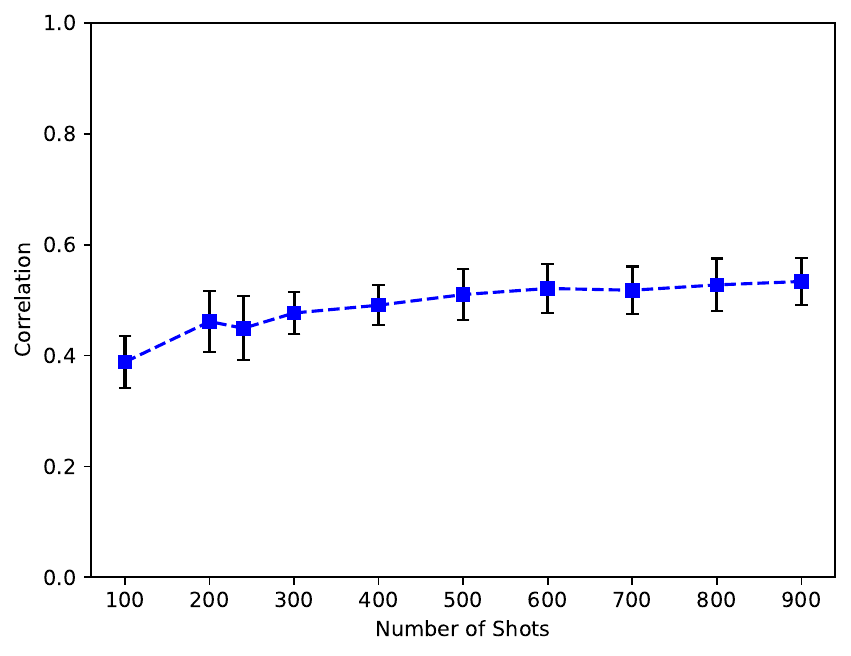}
        \caption{}
        \label{fig:Aquila_correlation_mean}
    \end{subfigure}
    \hspace{0.05\textwidth}
    \begin{subfigure}[b]{0.2\textwidth}
        \centering
        \includegraphics[width=\textwidth]{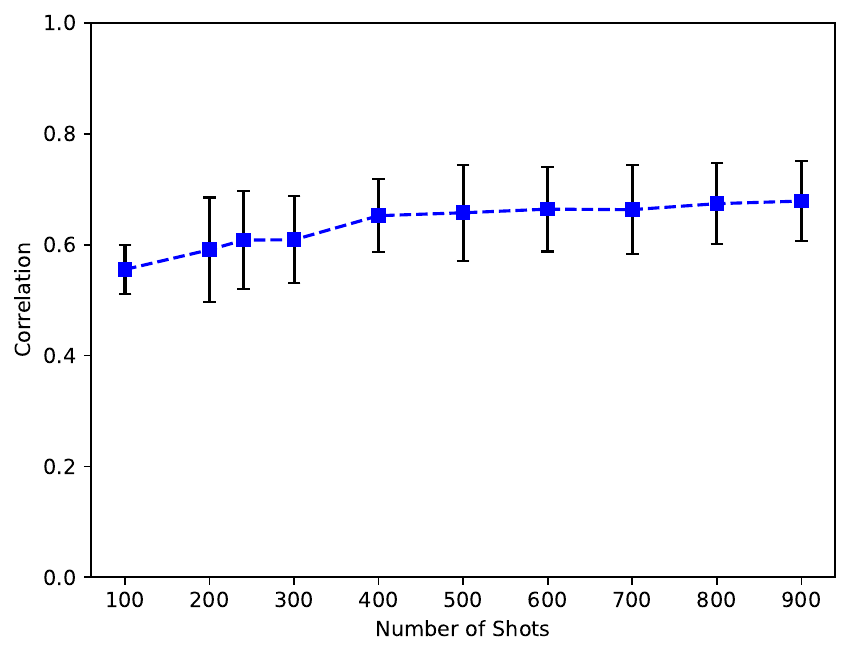}
        \caption{}
        \label{fig:Aquila_correlation_timestep_0}
    \end{subfigure}
    \hspace{0.05\textwidth}
    \begin{subfigure}[b]{0.2\textwidth}
        \centering
        \includegraphics[width=\textwidth]{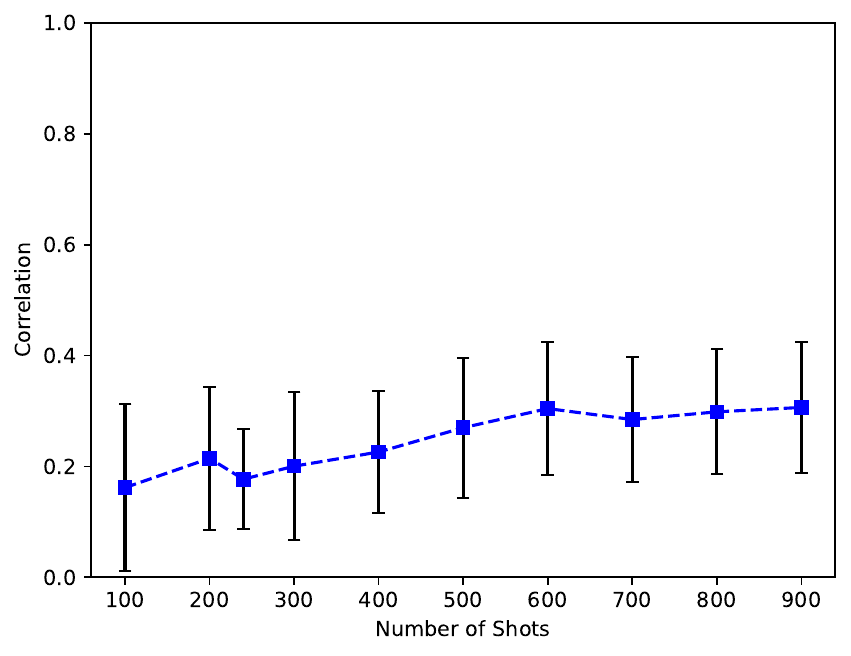}
        \caption{}
        \label{fig:Aquila_correlation_timestep_4}
    \end{subfigure}
    \caption{Statistical correlation between \textit{Aquila}'s QRC data (detuning encoding) and shot-based emulation as a function of number of shots and timestep, for 10 randomly chosen samples (a). Mean correlation across timesteps (b), as well as for the first (c) and last (d) timesteps are reported too.}
    \label{fig:correlation}
\end{figure}

To assess the quality of quantum simulation on \textit{Aquila} relative to the shot-based emulation, we analyzed the statistical correlation of the resulting QRC feature vectors, as shown in Fig.~\ref{fig:correlation} across different number of measurements and timesteps.
Under the assumption that \textit{Aquila}'s QRC features provide similar approximations of the noiseless features, despite the encoding, the shot measurement analysis was performed on $10$ samples randomly selected from the \texttt{CARDS\_2500} dataset using the detuning encoding. This analysis required running quantum simulations with up to $1000$ measurements per timestep on each sample. As observed, the correlation degrades at later timesteps, likely due to the onset of decoherence effects. Furthermore, increasing the number of measurements, particularly beyond $200$, does not lead to a substantial improvement in statistical correlation. Therefore, considering $N_{\text{repl}} = 6$, we selected an optimal number of measurement samples $n_{\text{opt}} = 240$, corresponding to $40$ shots per timestep and dataset sample. 

\begin{figure*}[htbp]
    \centering
    \includegraphics[width=0.95\textwidth]{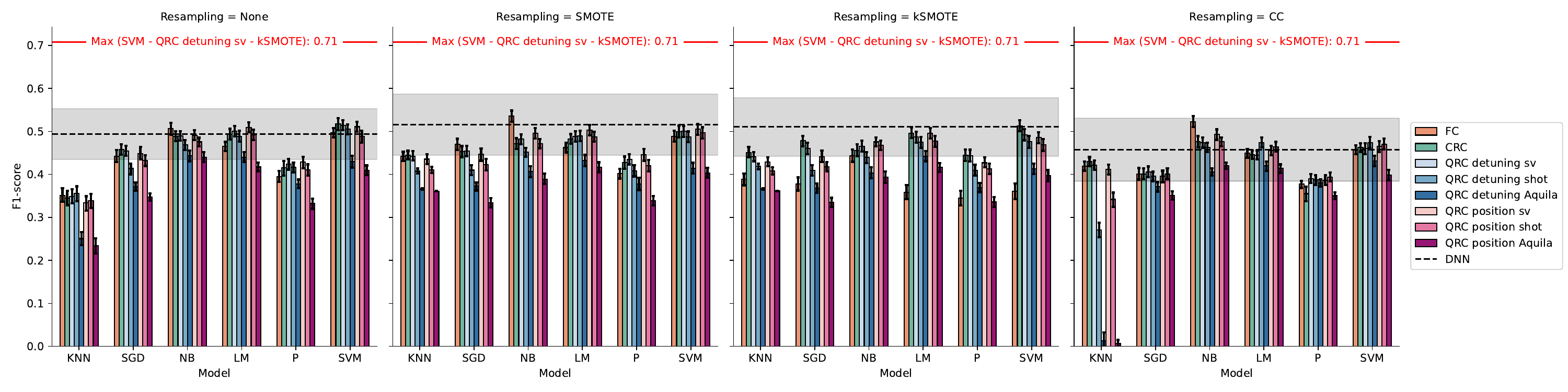}
    \caption{Benchmark of the classification results for the \texttt{CARDS\_1000}  dataset which compares real and emulated QRC, both detuning and position encoded, CRC and FC methods against the DNN. Dark shaded area shows the standard deviation for the DNN, while the red line indicates the maximum performance reached by any of the considered classifier.}
    \label{fig:1ksmall_results}
\end{figure*}

As reported in Section~\ref{sec:methodology}, we used a reduced dataset, \texttt{CARDS\_1000}, to compare the two embedding methodologies, before moving to a larger experimental test bench. Fig.~\ref{fig:1ksmall_results} reports the results of the classification procedure, in terms of F1-score, for this first dataset, considering the models and resampling methods mentioned in Section~\ref{sec:methodology}. The negative impact of hardware noise on classification performance is consistently observable: emulation results are always better than those obtained using the actual hardware. Comparing data from \textit{Aquila}, it can be noted that the detuning encoding generally performs better than or at least comparable to the position encoding in all the configurations considered. This is probably due to the larger shot-to-shot fluctuations of the atom positioning than the site-to-site fluctuations of the local detuning amplitude~\cite{wurtz2023aquilaqueras256qubitneutralatom, kornjača2024largescalequantumreservoirlearning}.

\begin{figure}[htbp]
    \centering
    \includegraphics[width=0.9\textwidth]{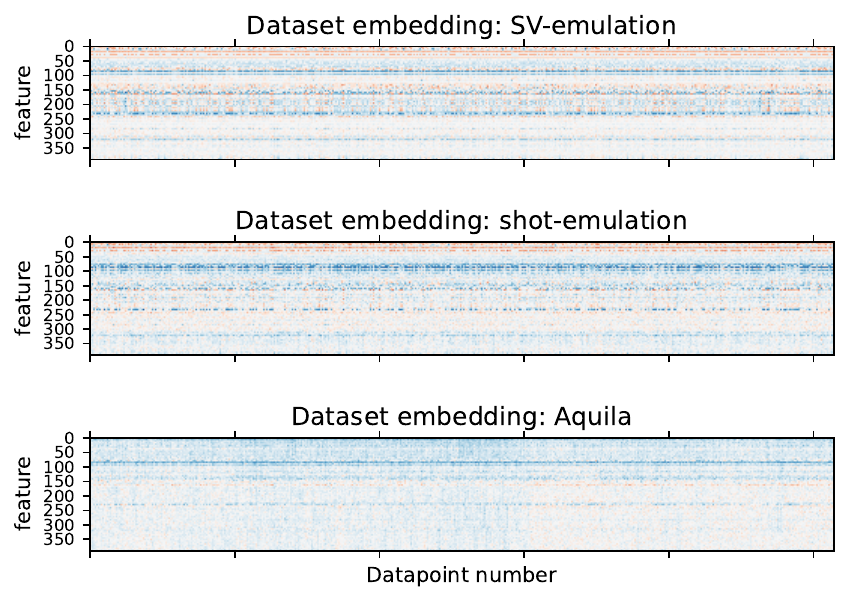}
    \caption{Comparison between QRC features of \texttt{CARDS\_2500} using real and emulated data.}
    \label{fig:embeddings}
\end{figure}

Based on this, a second round of experimental runs have been performed using \texttt{CARDS\_2500} encoded on the detuning pulse. 
The resulting QRC features computed using data from \textit{Aquila}, compared to the relative statevector and shot-based emulation, are shown in Fig.~\ref{fig:embeddings}. 

\begin{figure}[htbp]
    \centering
    \begin{subfigure}[b]{0.30\textwidth}
        \centering
        \includegraphics[width=\textwidth]{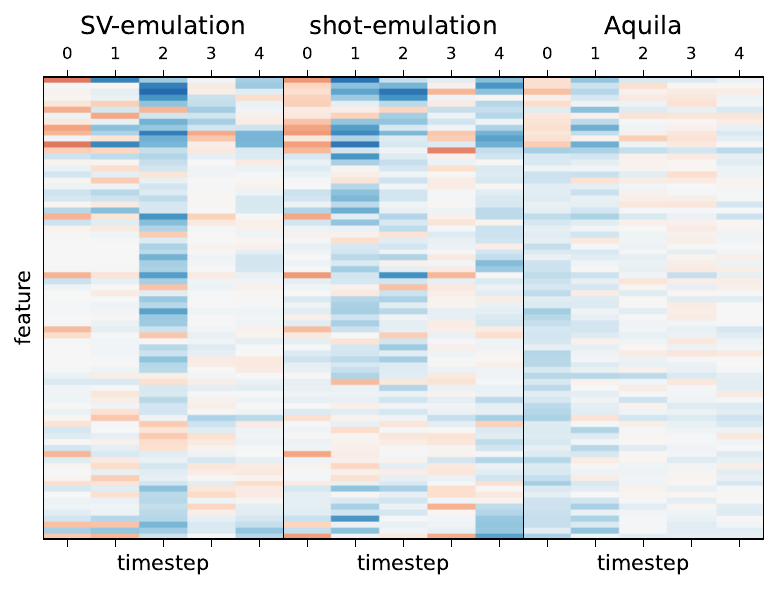}
        \caption{}
        \label{fig:embeddings_timestep_worst}
    \end{subfigure}
    \begin{subfigure}[b]{0.30\textwidth}
        \centering
        \includegraphics[width=\textwidth]{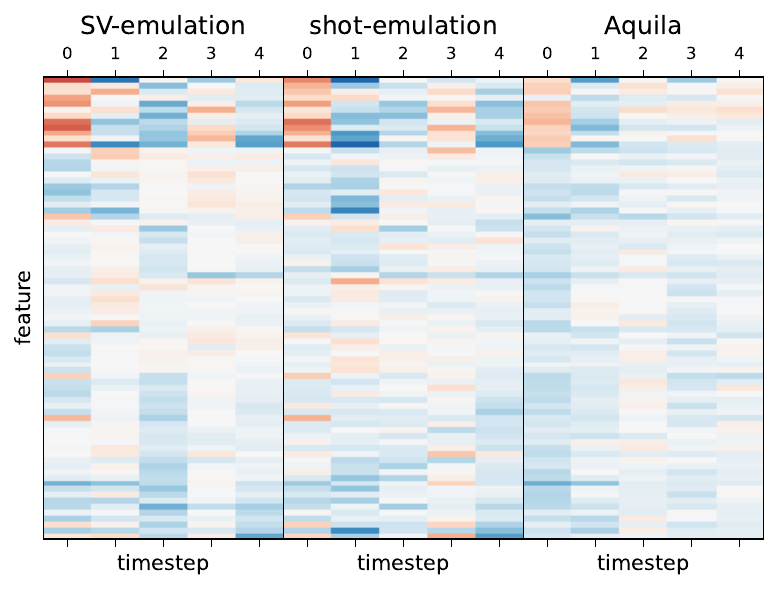}
        \caption{}
        \label{fig:embeddings_timestep_mean}
    \end{subfigure}
    \begin{subfigure}[b]{0.30\textwidth}
        \centering
        \includegraphics[width=\textwidth]{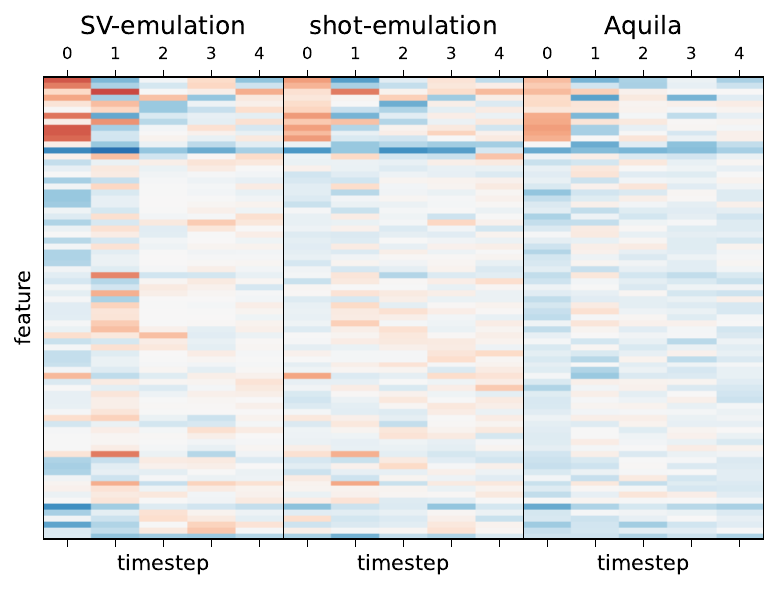}
        \caption{}
        \label{fig:embeddings_timestep_best}
    \end{subfigure}
    \caption{Comparison between QRC features (detuning encoded) using real and emulated data, segmented horizontally by timestep. Three samples with different correlation between real data and statevector emulation are reported: lowest (a), closest to mean (b), highest (c).}
    \label{fig:embeddings_timestep}
\end{figure}

To better visualize the section for each different timestep, the same comparison for the QRC features of a single sample of the dataset is reported in Fig.~\ref{fig:embeddings_timestep}. It is possible to notice the detrimental effect of the statistical noise and, more importantly, of the decoherence when using real data from the quantum simulator, which is already evident at the third timestep.

\begin{figure*}[htbp]
    \centering
    \includegraphics[width=0.95\textwidth]{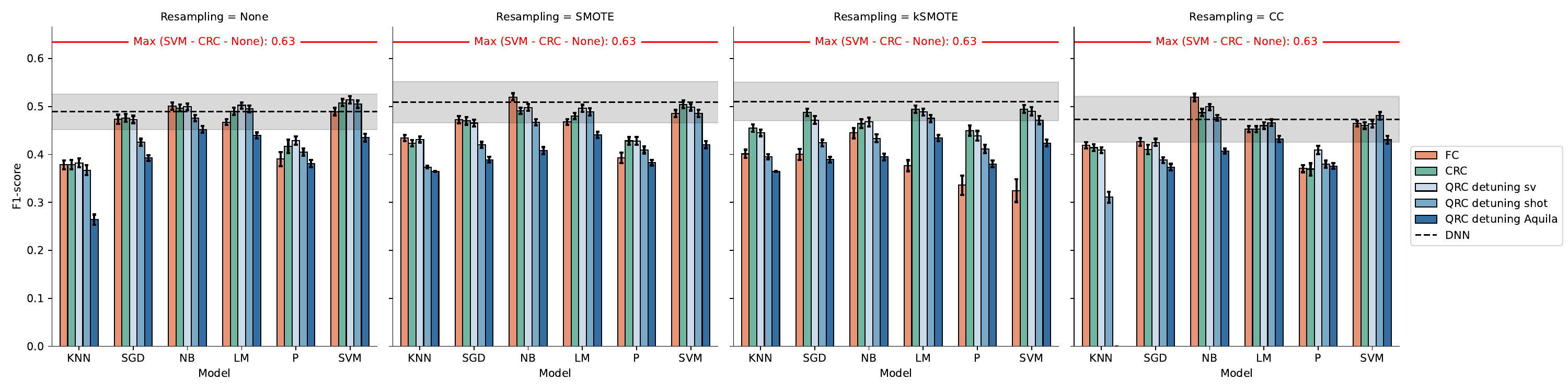}
    \caption{Benchmark of the classification results for \texttt{CARDS\_2500} which compares real and emulated QRC (detuning encoded), CRC and FC approaches against the DNN. Dark shaded area shows the standard deviation for the DNN, while the red line indicates the maximum performance reached by any of the considered classifier.}
    \label{fig:small_dataset_results}
\end{figure*}

The results of the classification on the \texttt{CARDS\_2500} are shown in Fig.~\ref{fig:small_dataset_results}.
In particular, noise still affects performance; however, the results obtained from the real hardware are generally closer to those from the emulated setup compared to the \texttt{CARDS\_1000} case. The main difference lies in the number of samples used, suggesting that the classifiers become more capable of recognizing and discarding machine-induced noise when more samples are available. This is an encouraging result for the use of QRC, compared to other QML techniques where the learning is performed only through the quantum module. This highlights the advantages of hybrid approaches that combine classical and quantum models.

\section{Conclusions}\label{sec:conclusions}

In this work, we have tested a Quantum Reservoir Computing (QRC) framework based on the use of neutral atoms trapped in optical lattices. The QRC is a promising approach for QML, as it allows for the processing of classical data using quantum systems. We have shown that the QRC can be implemented using a simple and efficient scheme, which can be easily adapted to different types of neutral atom platforms.

We have also demonstrated the potential of the QRC for financial classification tasks, in particular for predicting the default of credit card owners. The results show that the combination of QRC and simple classifiers can achieve a performance comparable to more complex neural networks, even with a small number of qubits and a limited amount of training data. This is particularly important for practical applications, where the availability of large datasets may be limited. 

Importantly, the considered classifiers require shorter training times and less computational resources than the deep neural network. This is a significant advantage. Given a classification problem and a performance threshold, QRC allows for a faster and more efficient training phase.

Our approach, thanks to the quantum hardware technology that we selected, can be potentially suitable for more complex ML tasks and applications. In fact, neutral atoms trapped in optical lattices allow for the implementation of large-scale quantum systems, with respect to other quantum computing platforms, because of its scalability and flexibility. In particular, for this technology, the register size is the main limiting factor for both the amount of input data that can be encoded in an atom array and the achievable sampling rate (through replication of such arrays). In this context, the trapping of thousand neutral atoms has already been demonstrated in a laboratory environment~\cite{PhysRevResearch.6.033104, PRXQuantum.5.030316}.

Further research directions may include applying gate-based implementations of QRC to the same dataset, enabling the use of different observables to construct the output feature vector. The digital version of QRC could allow for methods specifically tailored to the dataset at hand, leveraging more general Hamiltonian representations.

\appendix
\section{Appendix}\label{sec:appendix_methodology}
\subsection{Data cleaning and preprocessing}\label{subsec:cleaning_preprocessing}

We consulted the dataset description provided by UC Irvine~\cite{default_of_credit_card_clients_350} and identified inconsistencies in the data. Specifically, the feature \textit{MARRIAGE} is documented to have values 1 = married, 2 = single, and 3 = others, yet 54 samples contained an undocumented value of 0. Similarly, for the feature \textit{EDUCATION}, the documented values are 1 = graduate school, 2 = university, 3 = high school, and 4 = others. However, we found 345 samples with undocumented values: 14 with a value of 0, 280 with a value of 5, and 51 with a value of 6. Given the relatively small proportion of these samples, we opted to remove them to maintain consistency, as this did not significantly affect the overall distribution.

Additionally, we observed unexpected values in the payment history features, \textit{PAY\_1} through \textit{PAY\_6}. These features are supposed to have values of -1 for on-time payments, and values from 1 to 9 indicating payment delays in months. However, the dataset included integer values ranging from -2 to 8, and even after adding +1 to align with the expected range, 10228 samples still held an undocumented value of 0. Due to the large volume of these cases (out of the original 30,000 data points), it was impractical to remove them.

Building on the cleaned dataset, we then proceeded with the preprocessing step, exploiting Python's \texttt{scikit-learn} library~\cite{scikit-learn} for its robustness and ease of use.

As we employed a PCA-based approach, the first step is scaling. We chose an absolute value scaler, so all numerical features are scaled to the range $[-1, +1]$. For categorical features, namely \textit{SEX} and \textit{MARRIAGE}, we used a dummy variable approach (one-hot encoding). We also chose to treat the \textit{EDUCATION} feature as a numerical variable to maintain its ordinal nature and reduce the number of one-hot-encoded features.

After scaling, we applied PCA to reduce the dimensionality while preserving as much information as possible. We iteratively selected the Principal Components (PCs) until each additional component increases the explained variance by less than 0.001, leading us to retain 11 PCs that capture the main variance patterns in the data. Subsequently, we combined the original features based on their PCA scores through agglomerative clustering, a bottom-up hierarchical clustering approach. This method starts by treating each feature as its own cluster and then iteratively merges the two closest clusters until a specified threshold is reached; in this case, we set the threshold to half the maximum pairwise distance among features in the principal component space. At each step, the process merges clusters based on proximity, progressively building a hierarchy of clusters that represent similar patterns in the PCA-reduced space. Following this approach, we identified 12 distinct feature clusters as reported in Table~\ref{tab:feature_clusters}.

\begin{table}[ht]
\centering
\caption{Feature Clusters from Agglomerative Clustering}
\label{tab:feature_clusters}
\begin{tabular}{|c|l|}
\hline
\textbf{Cluster} & \textbf{Features} \\ \hline
Cluster 0 & \textit{PAY\_3}, \textit{PAY\_4} \\ \hline
Cluster 1 & \textit{PAY\_5}, \textit{PAY\_6} \\ \hline
Cluster 2 & \textit{PAY\_AMT1}, \dots, \textit{PAY\_AMT6} \\ \hline
Cluster 3 & \textit{MARRIAGE\_1} \\ \hline
Cluster 4 & \textit{BILL\_AMT1}, \dots, \textit{BILL\_AMT6} \\ \hline
Cluster 5 & \textit{MARRIAGE\_2} \\ \hline
Cluster 6 & \textit{PAY\_1} \\ \hline
Cluster 7 & \textit{LIMIT\_BAL} \\ \hline
Cluster 8 & \textit{AGE} \\ \hline
Cluster 9 & \textit{EDUCATION} \\ \hline
Cluster 10 & \textit{PAY\_2} \\ \hline
Cluster 11 & \textit{SEX\_1} \\ \hline
\end{tabular}
\end{table}

Finally, the scaled features within each cluster were aggregated into a single feature by averaging, effectively reducing redundancy and creating a compact feature set optimized for downstream tasks.

\subsection{Data resampling}
\label{apx:resampling}

Class imbalance is a common challenge in machine learning, often leading to biased predictions that favor the majority class. In the case of credit card default prediction, the \texttt{CARDS\_30000} dataset exhibits a significant imbalance, with 6605 instances labeled as defaulters (class 1) out of a total of 29601 samples. Although collecting additional real-world data would be ideal, in cases where that is not possible, resampling techniques such as oversampling and undersampling provide alternative strategies to balance the dataset and improve classification performance.

To allow the models to effectively learn from both classes, different resampling techniques were applied to the training set, while keeping the test set (4441 samples, 991 from class 1) and the validation set (4440 samples, 991 from class 1) unchanged. The distributions of the original and resampled datasets are summarized in Table~\ref{tab:all_dataset}.  

The efficacy of the resampling methods was also tested on the smaller datasets, namely \texttt{CARDS\_2500} and \texttt{CARDS\_1000}, with their size constrained by the availability of quantum resources. In these cases, resampling techniques were also used for the training set (see Tables~\ref{tab:2500_dataset} and~\ref{tab:1000_dataset}). For \texttt{CARDS\_2500}, the validation and test sets both consisted of 386 samples (86 of class 1) ; whilst, for \texttt{CARDS\_1000}, they are made of 150 samples (34 of class 1).

For the implementation of resampling methods, we utilized the \texttt{imblearn} (imbalanced-learn) Python library~\cite{lemaavztre2017imbalanced}.\\

\textbf{Oversampling Techniques}

Oversampling artificially increases the representation of the minority class to match the majority class distribution. In our study, we tested two variants of the \textit{Synthetic Minority Over-sampling Technique} (SMOTE)~\cite{chawla2002smote}.  

SMOTE generates synthetic samples by interpolating between a minority class data point and its nearest neighbors. Given a sample \( x \) from class 1 and one of its \( k \)-nearest neighbors \( x_k \), a synthetic sample \( x_{\text{new}} \) is generated as $ x_{\text{new}} = x + \lambda (x_k - x), \ \lambda \sim \mathcal{U}(0,1)$, 
where \( \mathcal{U}(0,1) \) denotes a uniform distribution; in our case, we set \( k = 5 \).

However, SMOTE has some weaknesses when dealing with imbalance and noise. Since it randomly selects a minority instance to oversample with uniform probability, densely populated minority areas are likely to be further inflated, while sparsely populated regions remain underrepresented. Another major concern is that SMOTE is susceptible to noise generation, as it does not distinguish between genuine and noisy minority samples. As a result, minority instances located among majority class instances may still be interpolated, potentially introducing unrealistic synthetic data.  

To address these issues, we tested a second variant: \textit{K-Means SMOTE} ($K$-SMOTE).  
K-SMOTE refines the synthetic sample generation process by first clustering minority class instances using $K$-Means before applying SMOTE within each cluster. This approach reduces the likelihood of generating unrealistic synthetic samples by preserving the natural distribution of class 1. In our setting $K = 2$, for the clustering step.\\

\textbf{Undersampling Techniques}

In scenarios where the majority class significantly outnumbers the minority class, undersampling can be an effective strategy to achieve a more balanced dataset. Our dataset contains a sufficient number of minority class samples, allowing us to explore undersampling methods. However, a key drawback of undersampling is the potential loss of valuable information from the majority class, which may lead to overly generalized patterns. To test such an approach, we employ the \textit{Cluster Centroids} (CC) method~\cite{lin2017clustering}.

The CC method applies $K$-Means clustering to identify representative centroids of class 0 and replaces multiple instances with their corresponding centroids. In this approach, $K$ is set to match the number of samples in class 1, ensuring a balanced dataset. This technique retains only the most informative samples from the majority class while preserving the overall feature space structure.

\begin{table}[htp]
    \centering
    \caption{\texttt{CARDS\_30000} dataset's training set class distributions with resampling methods.}
    \label{tab:all_dataset}
    \begin{tabular}{|l|c|c|c|}
        \hline
        \textbf{Resampling Method} & \textbf{Training Set} & \textbf{Class 1} & \textbf{Class 0} \\
        \hline
        None & 20720 & 4623 & 16097 \\
        SMOTE & 32194 & 16097 & 16097 \\
        K-SMOTE & 32198 & 16101 & 16097 \\
        Cluster Centroids & 9246 & 4623 & 4623 \\
        \hline
    \end{tabular}
\vspace{0.5cm}
    \centering
    \caption{\texttt{CARDS\_2500} dataset's training set class distributions with resampling methods.}
    \label{tab:2500_dataset}
    \begin{tabular}{|l|c|c|c|}
        \hline
        \textbf{Resampling Method} & \textbf{Training Set Size} & \textbf{Class 1} & \textbf{Class 0} \\
        \hline
        None & 1799  & 402 & 1397 \\
        SMOTE & 2794 & 1397 & 1397 \\
        K-SMOTE & 2797 & 1400 &  1397 \\
        Cluster Centroids & 804 & 402 & 402 \\
        \hline
    \end{tabular}
\vspace{0.5cm}
    \centering
    \caption{\texttt{CARDS\_1000} dataset's training set class distributions with resampling methods.}
    \label{tab:1000_dataset}
    \begin{tabular}{|l|c|c|c|}
        \hline
        \textbf{Resampling Method} & \textbf{Training Set Size} & \textbf{Class 1} & \textbf{Class 0} \\
        \hline
        None & 700  & 156 & 544 \\
        SMOTE & 1088 & 544 & 544 \\
        K-SMOTE & 1091 & 547 &  544 \\
        Cluster Centroids & 312 & 156 & 156 \\
        \hline
    \end{tabular}

\end{table}

\subsection{Classifiers}

\label{apx:classifiers}

This Section provides a description of the classification algorithms used in the study. For each algorithm, various hyperparameters were tuned to identify the optimal configuration based on performance on the validation set. Once the best-performing hyperparameters were determined, the corresponding model was trained and used to predict class labels on the test data. In addition, different resampling techniques, as introduced in Appendix~\ref{apx:resampling}, were evaluated in combination with the classification models.

Hyperparameter selection was guided by the F1-score, as accuracy can be misleading in imbalanced datasets, often yielding high values even when the minority class is not correctly classified. The classifiers were chosen based on their ease of training, aligning with the principles of reservoir computing. All model implementations rely on the \texttt{scikit-learn} Python library~\cite{scikit-learn}. Below, a detailed description of the models and the tested hyperparameter configurations is provided.\\

\textbf{Support Vector Machine}

Support Vector Machine (SVM) is a parametric linear classification algorithm that aims to separate two classes using a hyperplane in the feature space. Once the hyperplane is defined, classification is performed based on which side of the hyperplane a test point lies. If multiple hyperplanes satisfy the separation criterion, the one maximizing the margin, i.e., the distance between the hyperplane and the closest data points (support vectors) from each class, is selected.  

In cases where perfect linear separation is not possible, a soft-margin approach is used, allowing some margin violations. This is controlled by the penalty parameter \( C > 0 \), which is added to the Hinge loss function to be minimized. A higher \( C \) penalizes misclassified points more strictly, leading to a potentially tighter decision boundary.  
The parameter \( C \) was explored over the values \( \{10^{-1}, 10^0, 10^1, 10^2\} \). Given the imbalance in the dataset, the penalty parameter was further adjusted by incorporating the \textit{class weight} to account for differences in class distribution. \\  

\textbf{Na\"ive Bayes}

In this model, we tested the Na\"ive Bayes (NB) method in its Gaussian version, which assumes that the features follow a normal (Gaussian) distribution. Given a dataset with features \(\mathbf{x} = (x_1, x_2, \dots, x_n)\) and class labels \( y \), the Gaussian Na\"ive Bayes classifier models the conditional probability of each feature as  

\begin{equation}
P(x_i \mid y) = \frac{1}{\sqrt{2\pi\sigma_y^2}} \exp \left( -\frac{(x_i - \mu_y)^2}{2\sigma_y^2} \right),
\end{equation}

where \( \mu_y \) and \( \sigma_y^2 \) are the mean and variance of the feature \( x_i \) for class \( y \), estimated from the training data. The classification decision is then made using Bayes' theorem under the assumption that features are conditionally independent given the class label.  

Since the Gaussian Na\"ive Bayes classifier does not require hyperparameter tuning, the training and validation sets were merged to provide a larger dataset for more accurate distribution estimation.\\

\textbf{Logistic Model}

The Logistic Model (LM) is a parametric, discriminative binary classification algorithm. Specifically, it assumes that the predictors are linked to the mean of the response variable through the logistic (sigmoid) function:  

\begin{equation}
P(y = 1 \mid \mathbf{x}) = \frac{1}{1 + e^{-(\mathbf{w}^T \mathbf{x} + b)}},
\end{equation}

where \( \mathbf{x} \) represents the feature vector, \( \mathbf{w} \) is the weight vector, and \( b \) is the bias term. The probability of the negative class is simply given by \( P(y = 0 \mid \mathbf{x}) = 1 - P(y = 1 \mid \mathbf{x}) \).  

The algorithm estimates the model parameters by maximizing the likelihood function, or alternatively the maximum a posteriori (MAP) estimate if a regularization term is included. Given independent and identically distributed (i.i.d.) samples, the optimization process seeks the maximum likelihood estimate (MLE) of the parameters \( \mathbf{w} \). Unlike ordinary least squares regression, this optimization problem does not have a closed-form solution and is instead solved numerically using iterative methods such as gradient descent or Newton’s method.  

To prevent overfitting, a regularization term is typically added, ensuring that the model coefficients do not reach excessively high values. The hyperparameter \( C \) controls the strength of the regularization term in the objective function, with smaller values enforcing stronger regularization.  

In this study, we tested \( C \) over the values \\ \(\{10^{-3}, 10^{-2}, 10^{-1}, 10^0, 10^1, 10^2\}\) and incorporated \textit{class weights} into the penalty term to address dataset imbalance.\\

\textbf{Perceptron}

The Perceptron is a linear classification algorithm that updates its model only when misclassifications occur. It follows an online learning approach, where the weights are iteratively adjusted based on incorrect predictions. The decision function is defined as:

\begin{equation}
y = \text{sign}(\mathbf{w}^T \mathbf{x} + b),
\end{equation}

where \( \mathbf{w} \) is the weight vector, \( \mathbf{x} \) is the input feature vector, and \( b \) is the bias term. If a sample is misclassified, the weights are updated according to:

\begin{equation}
\mathbf{w} \leftarrow \mathbf{w} + \eta \cdot y \cdot \mathbf{x},
\end{equation}

where \( \eta \) is the learning rate.  

In this implementation, the Perceptron was trained using Stochastic Gradient Descent (SGD) without regularization (i.e., no penalty term). The model was tested with different learning rates: $\eta \in \{10^{-1}, 10^{-2}, 10^{-3}\}$

The training process ran for a fixed number of \( 1000 \) epochs, without early stopping. Additionally, class weights were incorporated to mitigate class imbalance. \\

\textbf{Stochastic Gradient Descent}

Stochastic Gradient Descent (SGD) is a simple yet highly efficient optimization approach for fitting linear classifiers and regressors under convex loss functions. 

In this implementation, the model was trained using the Hinge loss function:

\begin{equation}
L(\mathbf{w}) = \sum_{i} \max(0, 1 - y_i (\mathbf{w}^T \mathbf{x}_i + b)) + \lambda ||\mathbf{w}||_2^2,
\end{equation}

where \( y_i \in \{-1, 1\} \) represents the true class label, and \( \lambda \) is the regularization parameter controlling the L2 penalty. The L2 regularization term prevents overfitting by discouraging large weight values.  

Unlike standard SVM, which solves a constrained optimization problem using quadratic programming, SGD optimizes the loss function iteratively by updating the weights based on randomly sampled training points. This makes it computationally efficient for large datasets.  

Compared to the Perceptron, which updates its weights only on misclassified samples, SGD updates its weights at every iteration, even for correctly classified points, by minimizing the Hinge loss. Additionally, unlike the Perceptron, SGD incorporates an explicit regularization term.  

For this model, class weights were considered to address dataset imbalance, and the number of training \textit{epochs} was explored in the range $\{10, 20, \dots, 100\}$. \\

\textbf{K-Nearest Neighbors}

The \textit{K}-Nearest Neighbors (KNN) algorithm classifies a query point based on a majority vote among its \( k \) nearest neighbors. The assigned class is the most frequent among the \( k \) closest points in the dataset.  

The choice of \( k \) is crucial: a small \( k \) captures local structures but is sensitive to noise, while a larger \( k \) smooths decision boundaries. In this study, \( k \) was selected from the range $\{2, 3, \dots, 15\}$.
The Euclidean distance (\( L_2 \) norm) in the features space was used as the similarity metric.

\section*{Acknowledgment}
The authors acknowledge the support of \textit{QuEra} for interfacing with \textit{Aquila} and the useful discussions, as well as \textit{Amazon Braket} and \textit{Pawsey Supercomputing Research Center} for the research credit grant and support.

\bibliographystyle{IEEEtran}
\bibliography{bib_qrc}

\end{document}